\documentclass[a4paper,11pt]{article}
\pdfoutput=1
\usepackage[utf8x]{inputenc}
\usepackage{jheppub}
\usepackage{epsf}
\usepackage{graphicx}	
\usepackage{dcolumn}	
\usepackage{bm}			
\usepackage{amssymb, amsmath}
\usepackage{wasysym}
\usepackage{lipsum, color}

\usepackage{bbm}		

\usepackage{slashed}
\usepackage{hyperref}
\hypersetup{colorlinks, citecolor=bluscuro, linkcolor=black, urlcolor=bluscuro}
\definecolor{rossos}{cmyk}{0,1,1,0.55}
\definecolor{bluscuro}{rgb}{0.15, 0.2, .85}
\definecolor{bluchiaro}{cmyk}{1,.3,0.,0.1}

\let\oldquote\quote
\renewcommand\quote{\scriptsize\oldquote}
\let\oldquotation\quotation
\renewcommand\quotation{\scriptsize\oldquotation}

\newcommand{\gsim}{\lower.7ex\hbox{$\;\stackrel{\textstyle>}{\sim}\;$}}
\newcommand{\lsim}{\lower.7ex\hbox{$\;\stackrel{\textstyle<}{\sim}\;$}}

\def\beq{\begin{equation}}
\def\eeq{\end{equation}}
\def\bea{\begin{eqnarray}}
\def\eea{\end{eqnarray}}
\def\bmat{\begin{pmatrix}}
	\def\emat{\end{pmatrix}}
\def\bei{\begin{itemize}}
	\def\eei{\end{itemize}}
\def\eq#1{Eq.\ (\ref{#1})}

\def\Im{\, {\rm Im}}

\newcommand{\nn}{\nonumber}
\newcommand{\Tr}{\text{Tr}}

\usepackage[usenames,dvipsnames,svgnames]{xcolor}
\usepackage[normalem]{ulem}

\makeatletter
\def\section{\@startsection {section}{1}{\z@}{-3.5ex plus -1ex minus
		-.2ex}{2.3ex plus .2ex}{\large\bf}}
\def\subsection{\@startsection{subsection}{2}{\z@}{-3.25ex plus -1ex
		minus -.2ex}{1.5ex plus .2ex}{\normalsize\bf}}
\makeatother
\makeatletter

\@addtoreset{equation}{section}

\makeatother

\begin{document}
\date{\today}


\title{\begin{center} The Hierarchion, a Relaxion Addressing \\ the Standard Model's Hierarchies\end{center}}


\author[a]{Oz Davidi}
\author[a,b]{Rick S. Gupta}
\author[a]{Gilad Perez}
\author[a,c,d]{Diego Redigolo}
\author[a]{Aviv Shalit}

\affiliation[a]{Department of Particle Physics and Astrophysics, Weizmann Institute of Science,\\
	Rehovot 7610001, Israel}
\affiliation[b]{Institute for Particle Physics Phenomenology, Department of Physics, Durham University,\\
	DH1 3LE, Durham, United Kingdom}
\affiliation[c]{Raymond and Beverly Sackler School of Physics and Astronomy, Tel-Aviv University,\\
	Tel-Aviv 69978, Israel}
\affiliation[d]{School of Natural Sciences, Institute for Advanced Study, Einstein Drive,\\
	Princeton, NJ 08540, USA}
\abstract{We present a mechanism that addresses the electroweak, the strong CP, and the flavor hierarchies of the Standard Model (including neutrino masses) in a unified way. The naturalness of the electroweak scale is solved together with the strong CP problem by the Nelson-Barr relaxion: the relaxion field is identified with the pseudo-Nambu-Goldstone boson of an abelian symmetry with no QCD anomaly. The Nelson-Barr sector generates the ``rolling'' potential and the relaxion vacuum expectation value at the stopping point is mapped to the Cabibbo-Kobayashi-Maskawa phase. An abelian symmetry accounts for the Standard Model's mass hierarchies and flavor textures through the Froggatt-Nielsen mechanism. We show how the ``backreaction'' potential of the relaxion can be induced by a sterile neutrino sector, without any extra state with electroweak quantum numbers. The same construction successfully explains neutrino masses and mixings. The  only light field in our construction is the relaxion, which we call the \emph{hierarchion} because it is essentially linked to our construct that accounts for all the Standard Model hierarchies. Given its interplay with flavor symmetries, the hierarchion can be probed in flavor-violating decays of the Standard Model fermions, motivating a further experimental effort in looking for new physics in rare decays of leptons and mesons.}

\maketitle

\setcounter{tocdepth}{3}

\section{Introduction}\label{intro}

The Standard Model~(SM) suffers from several hierarchy issues: it cannot account for the very small scale associated with neutrino masses; it does not explain the structure of charged lepton masses, quark masses and mixing angles; nor does it provide a reason for the strong CP phase being at least ten orders of magnitude smaller than the Cabibbo-Kobayashi-Maskawa~(CKM) one. Last but not least, it also fails to explain why the Higgs mass is so much smaller than the Planck scale.

In this paper, we show how addressing the Higgs hierarchy problem with cosmological relaxation~\cite{Graham:2015cka} offers the possibility of addressing all the above shortcomings of the SM by a common mechanism - the spontaneous breaking of global symmetries. In our construction, the flavor textures and the hierarchical CP phases of the SM emerge from the spontaneous breaking of CP and flavor symmetries at a high scale, whereas the naturalness problem is addressed by the cosmological evolution of a single pseudo-Nambu-Goldstone boson~(pNGB), dubbed as the {\emph{hierarchion}}.

In cosmological relaxation models, the Higgs mass depends on a scalar field, the relaxion, whose dynamics is controlled by a non-generic potential~\cite{Gupta:2015uea}. Such a potential arises from the explicit breaking of a spontaneously broken abelian symmetry, $U(1)_{\text{clock}}$, which acts as a shift on the relaxion field. This symmetry is broken by two sequesterd sectors: the ``rolling'' sector and the ``backreaction'' sector. The two sectors have exponentially hierarchical charges under the $U(1)_{\text{clock}}$, resulting in exponentially hierarchical periodicities in the relaxion potential. A calculable setup with these features has been presented in Refs.~\cite{Choi:2014rja,Choi:2015fiu,Kaplan:2015fuy}. During inflation, the relaxion rolls down the rolling potential (with a very large periodicity), loses energy through Hubble friction, and eventually stops at the ``wiggles'' induced by the backreaction potential (with a smaller periodicity).

Our starting point is the Nelson-Barr relaxion of Ref.~\cite{Davidi:2017gir} which provides a unified solution to the Higgs hierarchy problem and the strong CP problem. We require CP to be an exact symmetry of the UV theory, broken only spontaneously by the hierarchion vacuum expectation value~(VEV). If the $U(1)_{\text{clock}}$ has no mixed anomaly with QCD, the hierarchion VEV does not induce a strong CP phase at tree-level, and can be mapped to the CKM phase \emph{\`a la} Nelson-Barr~(NB)~\cite{Nelson:1983zb, Barr:1984qx, Barr:1984fh}. To sequester the CP phase from the QCD sector, the up and/or the down sectors of the SM need to be extended. The relaxion rolling potential is hence generated by integrating out the extended quark sector.

The next step of our construction is to identify $U(1)_{\text{clock}}$ with a horizontal symmetry acting on the SM fermions. The hierarchy between the charged fermion masses is explained \emph{\`a la} Froggatt-Nielsen~(FN)~\cite{Froggatt:1978nt}, while the neutrino mass matrix is anarchic~\cite{Hall:1999sn,Haba:2000be,deGouvea:2003xe,GonzalezGarcia:2002dz}. In the quark sector, the main challenge we need to overcome is to fit the CKM structure, and keep the $U(1)_{\text{clock}}$ anomaly free at the same time.

The neutrino masses could be simply generated by adding sterile neutrinos that have a FN preserving Majorana mass (see for example Ref.~\cite{Ema:2016ops}). We present instead a sterile neutrino sector which also softly breaks the hierarchion shift symmetry, radiatively generating the backreaction potential. This construction gives a completely novel way of generating the relaxion wiggles, which does not involve \emph{any} new electroweak~(EW) charged state below the TeV scale. The smallness of neutrino masses is explained by a combination of seesaw mechanism and FN suppression. Moreover, the hierarchion mass gets interestingly connected to the scale of the SM neutrinos.  

\begin{figure*}[!t]
\centering
\includegraphics[scale=0.55]{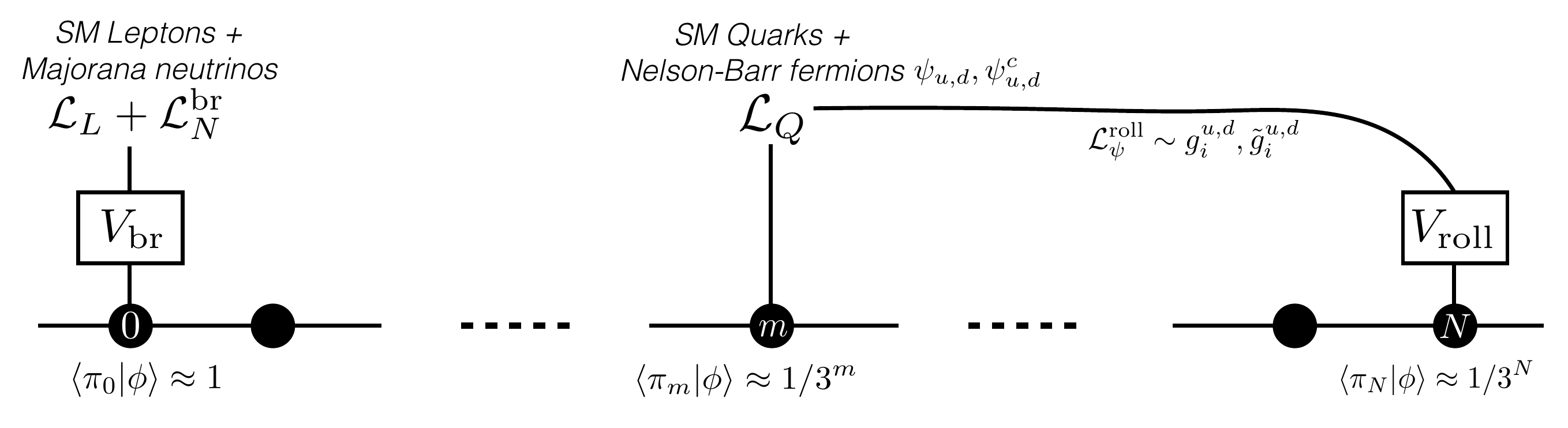}
\caption{Cartoon of the hierarchion construction. The Standard Model lepton sector, $\mathcal{L}_L$ in Eq.~\eqref{eq:fn_lep}, is connected to the $0^\textrm{th}$ site through higher dimension operators \emph{\`a la} Froggatt-Nielsen. On the same site, the sterile neutrino sector, $\mathcal{L}_N^{\text{br}}$ in Eq.~\eqref{fnb}, gives Majorana neutrino masses, and the ``backreaction'' potential for the hierarchion, $V_{\text{br}}$ in Eq.~\eqref{br}. The quark sector, $\mathcal{L}_Q$ in Eq.~\eqref{eq:fn_quark}, can also be connected \emph{\`a la} Froggatt-Nielsen to the $m^\textrm{th}$ site, where $0\leq m < N\,$. Two pairs of vector-like Weyl fermions, $(\psi_u, \psi^c_u)$ and $(\psi_d, \psi^c_d)$, are added to the $m^\textrm{th}$ site, and connected to the $N^\textrm{th}$ site through $\mathcal{L}_{\psi}^{\textrm{roll}}$ in Eq.~\eqref{eq:nbarr}. The portal in $\mathcal{L}_{\psi}^{\textrm{roll}}$ is controlled by small couplings, $g^{u,d}_{i}$ and $\tilde{g}^{u,d}_{i}$, and provides both the ``rolling'' potential, $V_{\text{roll}}$ in Eq.~\eqref{mu1}, and the Nelson-Barr phase.\label{fig:cartoon}}
\end{figure*}

This completes the dynamical picture of the hierarchion, which is at the same time the relaxion, the familon of a global horizontal symmetry~\cite{Wilczek:1982rv,Reiss:1982sq,Davidson:1984ik,Berezhiani:1990wn,Berezhiani:1990jj}, and the CKM phase of NB models. In Fig.~\ref{fig:cartoon} we present a cartoon of our construction, which we are going to detail in what follows. The parameter space of the hierarchion is presented in Fig.~\ref{fig:bounds}, both in the case where the backreaction is induced by the sterile neutrino sector or by a different \emph{ad hoc} sector, like in Ref.~\cite{Gupta:2015uea}.

In the case of sterile neutrino backreaction, the phenomenology of the hierarchion is completely dominated by its familon nature, which induces flavor-violating~(FV) decays of SM leptons and quarks~\cite{Feng:1997tn,Albrecht:2010xh}. In this regard, the hierarchion shares some similarities with axion models where the axion has FV couplings to fermions~\cite{Babu:1992cu,Ahn:2014gva,Celis:2014iua,Calibbi:2016hwq,Ema:2016ops,Marques-Tavares:2018cwm}. One crucial difference is that for the hierarchion with sterile neutrino backreaction, the couplings to the SM leptons are unsuppressed while the one to SM quarks are suppressed by a very small charge with respect to $U(1)_{\text{clock}}$. In this case, the hierarchion can be probed by looking at FV decays of the muon or of the tau, accompanied by missing energy~\cite{Bolton:1986tv,PhysRevD.38.2077,Jodidio:1986mz,Albrecht:1995ht,Igonkina:1996ar,Yoshinobu:2017jti}. While the bounds on $\tau\to \mu\, \phi$ are likely to be improved at Belle~II, the current best bound on $\mu\to e\, \phi$ is set by 30 year old data collected by the TRIUMF experiment~\cite{Jodidio:1986mz}. This bound can be only marginally improved by $\mu\to e\gamma$ experiments, like MEG~\cite{TheMEG:2016wtm}, MEG~II~\cite{Baldini:2018nnn} and Mu3e~\cite{Perrevoort:2016nuv}, looking at $\mu\to e\gamma\phi$ events (see also Ref.~\cite{Calibbi:2017uvl} for a comprehensive review of these experiments). The improvement could be substantial if a dedicated analysis on electron-only events is developed along the lines of Ref.~\cite{Jodidio:1986mz}. The latter is likely to require an upgrade of the detector and data-acquisition system (see for example Ref.~\cite{talk}).

If the backreaction sector is generic, the available parameter space gets bigger, and complementary constraints for hierarchion masses heavier than 1 keV can be derived from the relaxion couplings induced by its mixing with the SM Higgs~\cite{Flacke:2016szy}. The bounds from FV Kaon decays $K\to \pi\, \phi$ from combined E787 and E949 data~\cite{Adler:2008zza}, and the FV B-meson bound $B\to K\, \phi$ from Babar~\cite{Lees:2013kla} could also possibly play a role if the SM quark charges under $U(1)_{\text{clock}}$ are unsuppressed with respect to the lepton ones. These bounds will be improved in the near future by NA62~\cite{Fantechi:2014hqa}, and possibly by KOTO~\cite{Tung:2016xtx} and Belle~II~\cite{Aushev:2010bq}. 

This paper is organized as follows, in Sec.~\ref{sec:Review} we set our notation by reviewing the relaxion and its implementation in the clockwork mechanism. In Sec.~\ref{sec:Quark Sector} we present the quark sector of our construction: we first review the NB relaxion in Sec.~\ref{sec:NBR Basic Setup}, and then extend the construction to a full FN model in Sec.~\ref{sec:Quarks Basic Setup}. In Sec.~\ref{sec:charges} we present a concrete set of charges addressing the flavor puzzle in the quark sector. In Sec.~\ref{sec:Leptons Sector} we discuss the lepton sector. In particular, in Sec.~\ref{sec:neutrinos} we show how the FN mechanism makes it possible to generate the backreaction potential with sterile neutrinos. We then discuss the neutrino and charged lepton texture in this setup. In Sec.~\ref{sec:pheno} we show the parameter space of the hierarchion, and discuss its phenomenology. In Appendix~\ref{app:quarks} we discuss the radiative stability of the NB relaxion once extended to a full FN model, and an example of a UV completion is given in Appendix~\ref{app:UV}. In Appendix~\ref{app:leptons} we give some details about the parametric of the backreaction potential induced by the sterile neutrinos. Finally, in Appendix~\ref{app:lepton_pheno} we present the current status of searches of FV decays in the lepton sector, and discuss future improvements.


\medskip

\section{The Relaxion-Clockwork Mechanism}\label{sec:Review}

In the following section, we give a brief review of the relaxion~\cite{Graham:2015cka} and the clockwork~\cite{Choi:2014rja,Choi:2015fiu,Kaplan:2015fuy} mechanisms to set up our notation. The relaxion can be identified as the pNGB of a spontaneously broken $U(1)_{\text{clock}}$ symmetry. The $U(1)_{\text{clock}}$ gets broken explicitly by two sectors that are sequestered from each other: the rolling sector and the backreaction sector. Each of these explicit breaking sources induces a potential for the relaxion field with periodicity proportional to the relaxion decay constant divided by the charges carried by the fields in each sector.

The rolling potential can be written as
\begin{align}
V_{\text{roll}}&=\mu^2(\phi) H^{\dagger}\!H+\lambda_{H} (H^{\dagger}\!H)^{2} - \Lambda_{\text{roll}}^{4}\cos\!\frac{\phi}{F} \,,\label{mu1}\\
\mu^2(\phi)&\equiv\kappa \Lambda^2_H-\Lambda^2_H \cos\!\frac{\phi}{F} \,,\label{mu2}
\end{align}
where $H$ is the SM Higgs doublet, $\phi$ is the relaxion, $\Lambda_H$ is the UV cut-off of the Higgs effective field theory, and $F$ is the periodicity of the rolling sector. Notice that $\kappa\lesssim1$ while the scale $\Lambda_{\text{roll}}$ accounts for the fact that the threshold contributing to the rolling potential in Eq.~\eqref{mu1} can be different from the one giving the relaxion coupling to the Higgs $\Lambda_H$ in Eq.~\eqref{mu2} (see Refs.~\cite{Gupta:2015uea,Espinosa:2015eda}). Sometimes, for convenience, we will write the formula in terms of $r_{\text{roll}}\equiv\Lambda_{\text{roll}}^2/\Lambda_{H}^2\,$. 

Starting from an initial field value $\phi \lesssim \phi_c\approx -\vert F \cos^{-1} \kappa\vert\,$, such that $\mu^2(\phi)$ is positive and the EW symmetry unbroken, the relaxion rolls down during inflation, dissipating energy through Hubble friction.\footnote{Another possible source of energy dissipation one should consider is particle production if non-zero couplings with the SM bath are switched on~\cite{Hook:2016mqo,Choi:2016kke,Tangarife:2017rgl,Tangarife:2017vnd}. However, this is always subdominant compared to the Hubble friction if the relaxion velocity remains small throughout its rolling.} Once $\phi\gtrsim \phi_c\,$, $\mu^2(\phi)$ becomes negative, and the SM Higgs doublet develops a VEV, spontaneously breaking the EW symmetry. In a unitary gauge we can write
\begin{equation}
H=\begin{pmatrix} 0\\ v(\phi)+\frac{h}{\sqrt{2}}\end{pmatrix}\,,\qquad v(\phi)=\frac{\vert\mu(\phi)\vert}{\sqrt{2\lambda_{H}}}\,.
\end{equation}

After EW symmetry breaking~(EWSB) occurs, the backreaction potential generates the wiggles making it possible for the relaxion to stop rolling. The backreaction potential can be parametrized as
\begin{equation}
V_{\text{br}}=-M_{\text{br}}^{2}H^{\dagger}\!H \cos\!\frac{\phi}{f}-r^2_{\text{br}} M_{\text{br}}^{4} \cos\!\frac{\phi}{f}\,,\label{br}
\end{equation}
where $f$ is the periodicity of the backreaction sector, and $M_{\text{br}}$ parametrizes the mass threshold controlling the backreaction potential. Notice that the term proportional to $r^{2}_{\textrm{br}}$ in Eq.~\eqref{br} generates wiggles which are independent of the EW VEV, and hence would exist even before EWSB. In order for the relaxion mechanism to select the EW scale, we should therefore require this Higgs-independent contribution to be smaller than the a Higgs-dependent one. This usually implies an upper bound on $M_{\text{br}}$:
\begin{equation}
M_{\text{br}}\lesssim \frac{v}{r_\textrm{br}}\lesssim 4\pi v\,,\label{eq:2loop}
\end{equation}
with the rightmost bound coming from closing the Higgs loop (as emphasized in Refs.~\cite{Flacke:2016szy,Gupta:2015uea,Espinosa:2015eda}). The same bound is often quoted in terms of $\Lambda_{\text{br}}\equiv\sqrt{v M_{\text{br}}}$ and reads $\Lambda_{\text{br}}<\sqrt{4\pi} v\,$. As we will show in an explicit example in Sec.~\ref{sec:neutrinos}, the upper bound in Eq.~\eqref{eq:2loop} becomes more stringent if the Higgs-dependent and Higgs-independent contributions to the backreaction potential are generated at the same loop order.

At the end of its rolling, the relaxion stops at a local minimum of the potential, $\phi_0=\langle \phi\rangle\,$, where
\begin{align}
&\quad \sin \frac{\phi_0}{f}\sim\sin \frac{\phi_0}{F}\sim\mathcal{O}(1)\,,\label{eq:order1}\\
&\quad \partial_{\phi} V=0\quad\Rightarrow\quad \frac{\Lambda_\text{H}}{\Lambda_{\text{br}}} \sim \frac{\Lambda_\text{roll}}{r_{\text{roll}}^{1/2}\Lambda_{\text{br}}} \sim \left(\frac{F}{r_{\text{roll}}^2f}\right)^{1/4} \,.\label{eq:deriv}
\end{align}
An arbitrary constant can always be added to the potential in order to tune the cosmological constant to its observed value at the end of the rolling.

The minimization condition in Eq.~\eqref{eq:deriv} relates the ratio between the Higgs bare mass, $\Lambda_H$, and the backreaction scale, $\Lambda_{\text{br}}$, to the ratio of the periodicities in the relaxion potential, $F/f$. The problem of achieving $\Lambda_H\gg \Lambda_{\text{br}}$ is then translated to $F\gg f\,$, which requires a large hierarchy of charges between the rolling sector and the backreaction sector~\cite{Gupta:2015uea}.

One way to create such a hierarchy of charges is the so-called clockwork mechanism~\cite{Choi:2014rja,Choi:2015fiu,Kaplan:2015fuy}. This construction introduces $N+1$ spontaneously broken abelian symmetries at different sites of a moose diagram, as shown in Fig.~\ref{fig:cartoon}.\footnote{This construction is subject to a mild fine tuning problem that is related to the fact that all the $U(1)$'s have to be spontaneously broken if no extra symmetries are invoked~\cite{Flacke:2016szy}.} The potential is
\begin{equation}
V_{\text{clock}}=\sum_{j=0}^{N}\left(-m^2_\textrm{clock}\vert\Phi_j\vert^2+g_{\textrm{clock}}^{2}\vert\Phi_j\vert^4\right)\,,\label{eq:clock_pot}
\end{equation}
where for simplicity we assumed that the masses and quartic-couplings at every site are equal, so that the resulting decay constants of the associated Nambu-Goldstone bosons are given by $f=m_\textrm{clock}/\sqrt{2}g_{\text{clock}}\,$. The potential above has a $U(1)^{N+1}$ symmetry. The different sites are connected by $\epsilon$-suppressed operators, breaking explicitly $N$ of the abelian symmetries
\begin{equation}
\Delta V_{\text{clock}}=-\sum_{j=0}^{N-1}\left(\epsilon\Phi_{j}^\dagger\Phi_{j+1}^3+\text{h.c.}\right)\,.\label{eq:clocklinks}
\end{equation}
Expanding all the scalars around their VEV's
\begin{equation}
\Phi_j= \frac{1}{\sqrt{2}}(f+\rho_j)U_j\,,\qquad U_j=\text{e}^{i\pi_j/f}\,,\label{eq:notation}
\end{equation}
and taking $|\epsilon | \ll g_{\text{clock}}^2 \sim 1$ so that the radial modes can be decoupled, we get the potential for the angular modes, $\pi_j$,
\begin{equation}
\Delta V_{\text{clock}}=-\frac{\epsilon f^4}{4}\sum_{j=0}^{N-1} \cos\left(\frac{3\pi_{j+1}-\pi_{j}}{f}\right)\,.\label{clockwork}
\end{equation}
This leaves a single massless Nambu-Goldstone boson whose wave function is exponentially peaked at the $0^\textrm{th}$ site
\begin{equation}
\phi=c_\phi(N) \sum_{j=0}^{N}\frac{1}{3^{j}}\pi_j\,,
\end{equation}
where for $N\gg1\,$, the normalization constant is $c_\phi(N)\approx \sqrt{8/9}\,$, which we take to be 1 in what follows. Notice that $\phi$ non-linearly realizes the spontaneously broken $U(1)_{\text{clock}}$ symmetry:
\begin{align}
U(1)_{\text{clock}}: \qquad	& \pi_j\to \pi_j+\frac{2\pi}{3^{j}} f\alpha\,, \\
& \phi\to \phi+ 2\pi f\alpha\,,\notag
\end{align}
with $\alpha\in\left[0,3^{N}\right]\,$. All the other pNGB's get a mass $m_i^2\approx \epsilon f^2\,$, and zero VEV. 

The overlap between the massless eigenstate, $\phi$, and the site $j$ is $\langle\pi_j |\phi\rangle\approx 1/3^{j}\,$. Introducing explicit breaking of the global $U(1)_{\text{clock}}$ at the site $j$ of the clockwork would generate a potential for $\phi$ with periodicity of order $3^j f$. Therefore, by putting the backreaction sector at the $0^\textrm{th}$ site and the rolling sector at the $N^\textrm{th}$ site, we achieve the desired hierarchy for the relaxion potential $F/f\approx 3^N\,$, which, through Eq.~\eqref{eq:deriv}, solves the hierarchy problem between $\Lambda_H$ and $\Lambda_{\text{br}}$.

Going back to Fig.~\ref{fig:cartoon}, one can already visualize the different elements of the clockwork-relaxion, and the role they are going to play in the hierarchion construction. The $U(1)_{\text{clock}}$ is identified with a FN flavor symmetry. A NB sector breaks the $U(1)_{\text{clock}}$ at the last site, and also maps the ${\cal O}(1)$ phase $\frac{\phi_0}{F}$ to the CKM phase while not generating a strong CP phase at tree-level. Finally, as we discuss in Sec.~\ref{sec:Leptons Sector}, the backreaction potential can be generated by a $10\textrm{ GeV}$ scale sterile neutrino sector that breaks $U(1)_{\text{clock}}$ at the first site of the clockwork chain.


\medskip

\section{The Quark Sector}\label{sec:Quark Sector}

In this section, we present the quark sector of the hierarchion model. We begin by reviewing the NB relaxion of Ref.~\cite{Davidi:2017gir}, slightly extending the original construction, and then proceed to unify it with the FN mechanism.


\medskip

\subsection{The Nelson-Barr Relaxion}\label{sec:NBR Basic Setup}

The NB relaxion unifies the NB solution of the strong CP problem~\cite{Nelson:1983zb, Barr:1984qx, Barr:1984fh} with the relaxion solution of the hierarchy problem. NB models assume that CP is a good symmetry in the UV, which gets broken only spontaneously at an intermediate scale. Whereas the spontaneous breaking of CP generates an ${\cal O}(1)$ CKM phase, the strong CP phase generated at the intermediate scale is smaller than the observed bound, $\bar{\theta}_{\rm QCD}\leq 10^{-10}$~\cite{Afach:2015sja}. Once such a boundary condition is set by a UV model, $\bar{\theta}_{\rm QCD}$ remains small under renormalization group flow because the running of $\bar{\theta}_{\rm QCD}$ due to the CKM phase is a seven loop effect in the SM~\cite{Ellis:1978hq}. The model of Ref.~\cite{Davidi:2017gir} marries cosmological relaxation to the NB mechanism by utilizing the fact that the relaxion spontaneously breaks CP by stopping at an ${\cal O}(1)$ value (modulo $2 \pi$) of both $\theta_0\equiv\frac{\phi_0}{f}$ and $\theta_N\equiv\frac{\phi_0}{F}\,$. For the NB relaxion, the ${\cal O}(1)$ spontaneous CP breaking becomes a blessing as the NB texture allows for the phase $\theta_N\equiv\frac{\phi_0}{F}$ to be mapped to the CKM phase while keeping contributions to $\bar{\theta}_{\textrm{QCD}}$ small enough. This must be contrasted with the QCD relaxion models where the relaxion is identified with the axion, and an ${\cal O}(1)$ value of $\bar{\theta}_{\textrm{QCD}} \sim \theta_0\equiv\frac{\phi_0}{f}$ results in the `Relaxion CP Problem'. The ingredients required for such a NB solution automatically generate a rolling potential for the relaxion, leading to a unified solution of the strong CP and hierarchy problems.

The minimal model, presented in Ref.~\cite{Davidi:2017gir}, is an adaptation of the the model of Ref.~\cite{Bento:1991ez}, with $\Phi_N$ identified with the field that spontaneously breaks CP. We present here a simple extension of the NB relaxion where both the up and the down quark sectors are extended by the addition of two pairs of heavy vector-like fermion pair, $\left(\psi_u,\psi^c_u\right)$ and $\left(\psi_d,\psi^c_d\right)$, which are coupled to the SM as
\begin{equation}
\mathcal{L}_{\text{NB}}=\left[g^{d}_{j}\Phi_{N}+\tilde{g}^{d}_{j}\Phi_{N}^{*}\right]\psi_d d_{j}^{c}+\left[g^{u}_{j}\Phi_{N}+\tilde{g}^{u}_{j}\Phi_{N}^{*}\right]\psi_u u_{j}^{c}+\mu^d \psi_d\psi_d^c +\mu^u \psi_u\psi_u^c+\text{h.c.}\,,\label{eq:NB}
\end{equation}
where all couplings can be chosen to be real because of the underlying CP symmetry. The structure of the Lagrangian in Eq.~\eqref{eq:NB} is enforced by a discrete $\mathbbm{Z}_{2}$ symmetry under which $\psi_{u,d}\,$, $\psi_{u,d}^{c}$ and $\Phi_{N}$ are charged. The Weyl fermions $\psi_{u,d}$ and $\psi_{u,d}^c$ are in the fundamental and anti-fundamental of $SU(3)_C$, and carry opposite hypercharges: $\pm2/3$ for the up-type and $\pm1/3$ for the down-type. The $U(1)_N$ gets explicitly broken by the interactions of $\Phi_N$, the breaking being controlled by the product of the couplings $g^u \tilde{g}^u$ and $g^d\tilde{g}^d$. 

We now show that the above Lagrangian leads to a successful NB mechanism along the lines of Ref.~\cite{Bento:1991ez}. Setting $\Phi_N$ to its VEV, we can define the only CP violating couplings,
\begin{equation}
B^u_k=\frac{f}{\sqrt{2}}\left(g^{u}_{k}e^{i\theta_{N}}+\tilde{g}^{u}_{k}e^{-i\theta_{N}}\right)\,,\quad B^d_k=\frac{f}{\sqrt{2}}\left(g^{d}_{k}e^{i\theta_{N}}+\tilde{g}^{d}_{k}e^{-i\theta_{N}}\right)\,.\label{eq:BN}
\end{equation}
The $4\times4$ mass matrices of the up and down quarks at tree-level are
\begin{equation}\label{eq:quark_mass_tree_level}
M^{u}=\left(\begin{array}{cc}
(\mu^u)_{1\times1} & \left(B^u\right)_{1\times3}\\
(0)_{3\times1} & \left(vY^u\right)_{3\times3}
\end{array}\right)\,,\quad
M^{d}=\left(\begin{array}{cc}
(\mu^d)_{1\times1} & \left(B^d\right)_{1\times3}\\
(0)_{3\times1} & \left(vY^d\right)_{3\times3}
\end{array}\right)\,,
\end{equation}
where, again, the zeros of the $3\times 1$ blocks are enforced by the $\mathbbm{Z}_{2}$ symmetry, forbidding $Q\tilde{H}\psi_{u}^{c}$ and $QH\psi_{d}^{c}$. The special structure of the above matrices ensures that
\begin{equation}\label{eq:arg}
\bar{\theta}_{\textrm{QCD}}^{\textrm{tree-level}}={\rm Arg}(\mu^u\cdot{\rm det}(vY^{u}))+{\rm Arg}(\mu^d\cdot{\rm det}(vY^d))=0\ .
\end{equation}
Since the  $\mathbbm{Z}_{2}$ symmetry is spontaneously broken by the VEV of $\Phi_N$, threshold corrections induced by higher dimensional operators like $\Phi_N Q\tilde{H}\psi_{u}^{c}$ and $\Phi_N QH\psi_{d}^{c}$ can spoil the above structure, and need to be small enough to satisfy the upper bound on the strong CP phase. We exhibit a power counting where all the threshold corrections are controlled by $g^u \tilde{g}^u$ and $g^d\tilde{g}^d$, so that the upper bound on $\bar{\theta}_{\textrm{QCD}}$ results in an upper bound on these dimensionless couplings. A careful treatment of the loop corrections to the NB setup has been presented in Ref.~\cite{Davidi:2017gir}. We will come back to this issue in Appendix~\ref{app:quarks} after the full hierarchion model is presented. 

Now, we show that while a phase in $\left(B^{u,d}\right)_{1\times3}$ does not contribute to $\bar{\theta}_{\textrm{QCD}}$ at tree-level, it does generate an ${\cal O}(1)$ $\delta_{\textrm{CKM}}$. Upon integrating out the vector-like pair, we find the effective $3\times3$ mass matrices squared of the up and down sectors in the SM:
\begin{equation}
\left[M^{u,d}_{\textrm{eff}}M^{u,d\dagger}_{\textrm{eff}}\right]_{ij}\sim v^2 Y^{u,d}_{ik} Y^{u,d}_{jk}-\frac{v^2Y^{u,d}_{ik}B^{u,d\ast}_{k}B^{u,d}_{\ell}Y^{u,d}_{j\ell}}{(\mu^{u,d})^{2}+B^{u,d}_{n}B^{u,d\ast}_{n}}\,,\label{iout}
\end{equation}
where we have used $(\mu^{u,d})^{2}+B^{u,d}_{n}B^{u,d\ast}_{n}\gg v^2Y^{u,d}_{ik}Y^{u,d}_{jk}$ for all $i$ and $j$. Recall that the diagonalizing matrices of the above mass matrices, $V_{u,d}^L$, form the CKM matrix $V_{u}^{L\dagger} V_{d}^L$. Assuming
\begin{gather}
\left|\vec{g}^{\,u,d}\times\vec{\tilde{g}}^{\,u,d}\right|/\left|\vec{g}^{\,u,d}+\vec{\tilde{g}}^{\,u,d}\right|^2\sim1\,,\label{eq:cond1}\\
\mu^{u,d} \lesssim \left|B^{u,d}_{k}\right|\,,\label{eq:cond2}
\end{gather}
there is an ${\cal O} (1)$ phase in the rightmost term of Eq.~\eqref{iout}, which translates to $\delta_{\rm CKM}\sim\mathcal{O}(1)\,$, and a Jarlskog invariant~\cite{Jarlskog:1985ht} that has the correct observed magnitude.

Interestingly, the NB construction also automatically generates the rolling potential for the relaxion. Indeed, the Lagrangian in Eq.~\eqref{eq:NB} clearly breaks the $U(1)_N$ symmetry at the last site, both in the up and in the down sector. Because of the bigger Yukawa couplings, the up sector dominates the radiative corrections to the rolling terms in the relaxion potential. By matching to Eq.~\eqref{mu1} and Eq.~\eqref{mu2}, we can then estimate the Higgs cut-off and the rolling potential scale as
\begin{align}
\Lambda_H^2&\sim\frac{g^{u}_{j} \tilde{g}^{u}_{k}\left((Y^{u})^{T}Y^u\right)_{jk}}{16 \pi^2} f^2\log\left[\frac{m_\textrm{clock}^2}{\left(\mu^{u}\right)^{2}}\right]\gtrsim\frac{y_t^2\left(\mu^{u}\right)^{2}}{16\pi^2}\log\left[ \frac{m_\textrm{clock}^2}{\left(\mu^{u}\right)^{2}}\right]\,,\label{eq:matchingH}\\
\Lambda_{\text{roll}}^2&\sim\frac{\sqrt{g^{u}_{j} \tilde{g}^{u}_{j}} g_{\text{clock}} f^2}{4\pi}\gtrsim \frac{\mu^u m_\textrm{clock}}{4\pi}\,,\label{eq:matchingnb}
\end{align}
where $g_{\textrm{clock}}$ and $m_\textrm{clock}$ are the ${\cal O}(1)$ clockwork quartic and mass term in Eq.~\eqref{eq:clock_pot}, and we used the NB condition in Eq.~\eqref{eq:cond2} to get the inequalities on the right hand side. It is worth emphasizing that the existence of the physical phase, $\theta_N$ in Eq.~\eqref{eq:NB}, which ultimately arises because $\pi_N$ takes a VEV, is closely related to the generation of the rolling potential. This is because in the absence of the rolling potential, $\pi_N$ has only derivative couplings to other SM fields, and no physical phase can arise in the quark mass matrix in such a scenario. The scaling of Eq.~\eqref{eq:matchingH} and of Eq.~\eqref{eq:matchingnb} are a consequence of the \emph{hard} breaking of the $U(1)_N$ symmetry. These imply
\begin{equation}
\frac{1}{r_{\text{roll}}}\equiv\frac{\Lambda_{H}^2}{\Lambda_{\text{roll}}^2}\lesssim 10^{-4}\cdot \left( \frac{\sqrt{g^u \tilde{g}^u}}{10^{-3}}\right)\,,\label{upper_bound}
\end{equation}
where in the last inequality, we substitute the upper bound on $g^u$ and $\tilde{g}^u$ that is required to keep the radiative corrections to the NB construction under control and will be derived later. This inequality, together with Eq.~\eqref{eq:deriv}, poses a serious limit on how much the NB relaxion can push up the Higgs UV threshold $\Lambda_H$ compared to other relaxion models where $\frac{1}{r_{\text{roll}}}\simeq 1\,$. In what follows, we show that besides this constraint, one can still get a large viable parameter space for the simplest possible NB relaxion, where also the flavor puzzle and neutrino masses are addressed in a coherent way. Another possible direction would be to relax the upper bound in Eq.~\eqref{upper_bound} by breaking the $U(1)_N$ only \emph{softly} while still generating an $\mathcal{O}(1)$ phase from the relaxion VEV. We leave this challenge for future investigations. 


\medskip

\subsection{Quark Sector: Basic Setup}\label{sec:Quarks Basic Setup}

We now embed the NB relaxion in a FN setup, by identifying $U(1)_{\text{clock}}$ with a horizontal flavor symmetry. Both the SM Yukawa couplings and the vector-like masses of the new fermion pairs of the NB relaxion in Eq.~\eqref{eq:NB} arise now as higher dimensional operators induced by the underlying FN construction.

The scalar $\Phi_m$ at the $m^\textrm{th}$ site of the clockwork chain is identified with the flavon field, which couples to SM quarks via non-renormalizable interactions, and $U_m= e^{i\pi_m/f}$ is the FN familon~\cite{Wilczek:1982rv}. The resulting effective Lagrangian is:\footnote{In addition to operators of the form $(\hat{\Phi}_{m}/\Lambda_{u,d})^{3^{n}}$, one can consider lower dimension operators of the form $\hat{\Phi}_{m-n}/\Lambda_{u,d}\,$, that are allowed by the $U(1)_\textrm{clock}$ symmetry. This can potentially jeopardize the FN suppression of the SM Yukawas. However, we assume that the SM quarks are \emph{localized} at the $m^\textrm{th}$ site, so that no tree-level interactions are allowed with clockwork fields other than $\hat{\Phi}_{m}$. Consequently, the dangerous operators above will be only radiatively generated and are further suppressed by powers of $\epsilon$, which make them subdominant. Clearly, this issue does not arise if $m=0\,$.}
\begin{align}\label{eq:fn_quark}
\mathcal{L}_{Q} & = y^{u}_{jk}\cdot\left(\frac{\hat{\Phi}_m}{\Lambda_u}\right)^{\left|\left[Q_j\right]+\left[u^c_k\right]\right|}Q_{j}\tilde{H}u^{c}_{k} + y^{\psi_{u}}\left(\frac{\hat{\Phi}_{m}}{\Lambda_u}\right)^{\left|\left[u_\psi\right]+\left[\psi^c_u\right]\right|-1}\hat{\Phi}_{m}\psi_{u}\psi_{u}^{c} \notag\\
& + y^{d}_{jk}\cdot\left(\frac{\hat{\Phi}_m}{\Lambda_d}\right)^{\left|\left[Q_j\right]+\left[d^c_k\right]\right|}Q_{j}Hd^{c}_{k} + y^{\psi_{d}}\left(\frac{\hat{\Phi}_{m}}{\Lambda_d}\right)^{\left|\left[d_\psi\right]+\left[\psi^c_d\right]\right|-1}\hat{\Phi}_{m}\psi_{d}\psi_{d}^{c} \notag\\
&\supseteq Y^{u}_{jk}\cdot U_m^{\left[Q_j\right]+\left[u^c_k\right]} Q_{j}\tilde{H}u^{c}_{k} + \mu^{u}U_{m}^{\left[\psi_{u}\right]+\left[\psi_{u}^{c}\right]}\psi_{u}\psi_{u}^{c}\notag\\
&+ Y^{d}_{jk}\cdot U_{m}^{\left[Q_j\right]+\left[d^c_k\right]} Q_{j}Hd^{c}_{k} + \mu^{d}U_{m}^{\left[\psi_{d}\right]+\left[\psi_{d}^{c}\right]}\psi_{d}\psi_{d}^{c}\,,
\end{align}
where $Q$ are the SM $SU(2)$ doublet quarks, and $d^c$ and $u^c$ are the SM $SU(2)$ singlet quarks. The charges of the fields presented above are normalized with respect to the $U(1)_m$ FN symmetry,\footnote{To obtain the charge of any field that is charged under the $U(1)_m$ symmetry, with respect to $U(1)_\text{clock}$, one must rescale its $U(1)_m$ charge by a factor of $3^{-m}$, namely $\left[X\right]_m=3^m\left[X\right]_{\textrm{clock}}\,$.} and the SM Higgs is uncharged without loss of generality, $[H]=0\,$. We normalize the charges such that $\left[\Phi_m\right]=-1\,$, and $\hat{\Phi}_m=\Phi_m/\Phi_m^*$ if the FN charges are positive/negative, so that the FN symmetry is respected. The cut-off scales, $\Lambda_u$ and $\Lambda_d$, would be associated to the masses of the FN messengers in explicit UV completions~\cite{Leurer:1992wg,Leurer:1993gy}. After $\Phi_m$ takes a VEV $f$ (see Eq.~\eqref{eq:notation}), powers of $\sqrt{2}\varepsilon_{u,d}=f/\Lambda_{u,d}<1$ account for the hierarchies in the Yukawa matrices:
\begin{equation}
Y^{u}_{jk}=y^{u}_{jk}\cdot\varepsilon_{u}^{\left|\left[Q_{j}\right]+\left[u^{c}_{k}\right]\right|}\cdot e^{i \left(\left[Q_{j}\right]+\left[u^{c}_{k}\right]\right) \theta_m}\,,\quad Y^{d}_{jk}=y^{d}_{jk}\cdot\varepsilon_{d}^{\left|\left[Q_{j}\right]+\left[d^{c}_{k}\right]\right|}\cdot e^{i \left(\left[Q_{j}\right]+\left[d^{c}_{k}\right]\right) \theta_m}\,,
\label{eq:yukquark}
\end{equation}
and the hierarchies between $\mu^{u, d}$ and $f$:
\begin{equation}
\mu^{u}= y^{\psi_{u}}\varepsilon_{u}^{\left|\left[\psi_{u}\right]+\left[\psi_{u}^{c}\right]\right|-1} e^{i\left(\left[\psi_{u}\right]+\left[\psi_{u}^{c}\right]\right)\theta_{m}}\frac{f}{\sqrt{2}}\,,\quad\mu^{d}=y^{\psi_{d}} \varepsilon_{d}^{\left|\left[\psi_{d}\right]+\left[\psi_{d}^{c}\right] \right|-1}e^{i\left(\left[\psi_{d}\right]+\left[\psi_{d}^{c}\right]\right)\theta_{m}}\frac{f}{\sqrt{2}}\,.
\label{eq:mus}
\end{equation}
Alternatively, if one chooses $\left[\psi_u\right]=-[\psi^c_u]$ and $\left[\psi_d\right]=-[\psi^c_d]\,$, the vector-like masses $\mu^{u,d}$ can be introduced directly as technically natural parameters. A $\mathbbm{Z}_{2}$ symmetry, under which $\psi_{u,d}\,$, $\psi_{u,d}^{c}$ and $\Phi_N$ are odd, prohibits the otherwise allowed gauge invariant operators, such as $\hat{\Phi}_m^{\left|\left[Q\right]+\left[\psi^c_u\right]\right|}Q\tilde{H}\psi^c_u$ or $\hat{\Phi}_m^{\left|\left[\psi_u\right]+\left[u^c\right]\right|}\psi_u u^c\,$, thus forbidding any Yukawa interaction between the vector-like quarks, $\psi_{u,d}$ and $\psi^c_{u,d}\,$, and the SM fermions.

An explicit breaking of the $U(1)_N$ is introduced here along the lines of Eq.~\eqref{eq:NB}. The leading operators that break the $U(1)_N$ symmetry but preserve both the $U(1)_m$ and $\mathbbm{Z}_{2}$ symmetries are

\begin{equation}
\mathcal{L}_{\psi}^{\textrm{roll}}=\left[g^{u}_{k}\Phi_{N}+\tilde{g}^{u}_{k}\Phi_{N}^{\ast}\right]\left(\frac{\hat{\Phi}_m}{\Lambda_u}\right)^{\left|\left[\psi_u\right]+\left[u^c_k\right]\right|} \psi_u\, u^{c}_{k}+\left[g^{d}_{k}\Phi_{N}+\tilde{g}^{d}_{k}\Phi_{N}^{\ast}\right]\psi_d\, d^{c}_{k}+\textrm{h.c.}\,.\label{eq:nbarr}
\end{equation}
In the above, we have assumed the same charge, $\left[d^c\right]=-\left[\psi_d\right]\,$, for all the SM down $SU(2)$ singlet quarks. This choice is consistent with the down-type mass spectrum of the SM, and is crucial in order to generate an ${\cal O}(1)$ $\delta_{\textrm{CKM}}$.  In Appendix~\ref{app:UV}, we show how Eq.~\eqref{eq:nbarr} is recovered after the heavy FN fields are integrated out at tree-level, where $\vec{g}^{\,u,d}$ and $\vec{\tilde{g}}^{\,u,d}$ arise as a linear combination of the explicit breaking couplings of the UV. Requiring $\bar{\theta}_{\textrm{QCD}}$ in Eq.~\eqref{eq:arg} to vanish gives us
\begin{equation}
n_{\text{QCD}}=\sum_{j=1}^3\left( 2 \left[Q_j\right] +\left[u^c_j\right] +\left[d^c_j\right]\right)+[\psi_u]+[\psi^c_u]+[\psi_d]+[\psi^c_d]=0\, .\label{eq:ABJzero}
\end{equation}
This corresponds to $U(1)_{\text{clock}}$ having a zero Adler-Bell-Jackiw~(ABJ) anomaly with QCD, $n_{\text{QCD}}$ being the anomaly coefficient. Alternatively, one can see that if Eq.~\eqref{eq:ABJzero} is satisfied, $\mathcal{L}_{Q}$ can be made real by quark chiral rotations, without inducing a $G\tilde{G}$ term.

Now, we turn to the condition in Eq.~\eqref{eq:cond1}, which is the crucial requirement to get an $\mathcal{O}(1)$ contribution to the CKM phase. We define again $\vec{B}^{u,d}$ to be the $1\times3$ blocks of the mass matrices, as in Eq.~\eqref{eq:quark_mass_tree_level}, so that elements of $\vec{B}^{u}$ now include powers of $\varepsilon_{u}$. The fact that the components of $\vec{B}^{u}$ have different degrees of $\varepsilon_u$ suppression means that this condition cannot be satisfied in the up sector, unless hierarchical $g^{u}_{k}$ and $\tilde{g}^{u}_{k}$ are introduced. However, introducing a flavor pattern in these couplings will be somehow distasteful. On the other hand, our choice of down FN charges $\left[d^c\right]=-\left[\psi_d\right]$ implies that the condition in Eq.~\eqref{eq:cond1} can be satisfied for $g^{d}_{k}\sim\tilde{g}^{d}_{k}\sim g_d\,$. This results in an $\mathcal{O}(1)$ contribution to $\delta_{\rm CKM}$ from the down sector as long as $\mu^d\lesssim \left|g_d\right| f\,$. From now on we will simplify our discussion by taking $g^{u,d}_{k}\sim\tilde{g}^{u,d}_{k}\sim g_{u,d}\,$, and assuming $\mu^{d}\sim\left|B^{d}_{k}\right|$ and $\mu^{u}\gtrsim \left|B^{u}_{k}\right|\,$. This can be achieved by appropriately choosing that charges of the vector-like fermion pairs in Eq.~\eqref{eq:mus}.

So far, we have admitted only the lowest order breaking of the $U(1)_N$ symmetry by the couplings $g^{u,d}_{k}\sim\tilde{g}^{u,d}_{k}\sim g_{u,d}$ in Eq.~\eqref{eq:nbarr}. Further higher order corrections, proportional to powers of $g_{u,d}\,$, can potentially spoil the structure of the quark mass matrices required for a successful NB mechanism. Such effects are either caused by radiative corrections at the scale $f$, or by irrelevant operators generated at the scale $\Lambda_{u,d}$. In the presence of the $U(1)_N$ breaking terms in Eq.~\eqref{eq:nbarr}, the states at the scale $\Lambda_{u,d}$ that generate the FN operators of Eq.~\eqref{eq:fn_quark}, also generate dangerous irrelevant operators. For instance, one dangerous operator is
\begin{equation}
\mathcal{L}\supset\frac{c_u}{16\pi^{2}}\left(\frac{\hat{\Phi}_{m}}{\Lambda_{u}}\right)^{\left|[Q_{j}]+ [\psi_{u}^{c}]\right|} \sum_{k}\left(\tilde{g}^{u}_{k}\frac{\Phi_{N}}{\Lambda_{u}}+g^{u}_{k}\frac{\Phi_{N}^{*}}{\Lambda_{u}}\right)Q_{j}\tilde{H}\psi_{u}^{c}\,,\label{eq:mostdanger}
\end{equation}
$c_u$ being an $\mathcal{O}(1)$ number. This operator gives a contribution to the $3\times 1$ block of the up mass matrix in Eq.~\eqref{eq:quark_mass_tree_level}, and thus to $\bar{\theta}_{\textrm{QCD}}$.

In Appendix~\ref{app:quarks}, we discuss the various contributions to $\bar{\theta}_{\textrm{QCD}}$. For a very general class of UV models, we find the largest contributions to be
\begin{equation}
\bar{\theta}_\textrm{QCD}\sim\sum_{q\in\left\{u,d\right\}}\frac{\alpha^{q}_{jk}}{16 \pi^2}\left(g^{q}_{j}\tilde{g}^{q}_{k}-g^{q}_{k}\tilde{g}^{q}_{j}\right)\frac{\left|\langle \Phi_{N}\rangle\right|^{2}}{\Lambda_{q}^{2}}\sin\left(2\theta_N\right)\,,
\end{equation}
where $\alpha^{q}_{jk}$ are ${\cal O}(1)$ numbers. The bound $\bar{\theta}_{\textrm{QCD}}\lesssim10^{-10}$ (from measurements of the neutron electric dipole moment~\cite{Afach:2015sja}) translates into an upper bound on the explicit breaking of the $U(1)_N$: 
\begin{equation}\label{eq:bound}
g_{u,d} \lesssim 10^{-3} \cdot \left(\frac{0.1}{\varepsilon_{u,d}}\right)\,.
\end{equation}
We further show in detail that once Eq.~\eqref{eq:bound} is fulfilled, the radiative threshold effects at the scale $f$, considered in Refs.~\cite{Bento:1991ez, Dine:2015jga}, are also under control. Notice that unlike Refs.~\cite{Bento:1991ez, Dine:2015jga}, we introduce $U(1)_N$ breaking in a controlled way by the couplings $g^{u,d}_{j}\sim\tilde{g}^{u,d}_k\sim g_{u,d}\,$. This allows us to estimate any higher order contribution to $\bar{\theta}_{\textrm{QCD}}$ parametrically in powers of $g_{u,d}$ by using symmetry arguments only.

As in the case of NB relaxion, the $U(1)_N$ breaking in Eq.~\eqref{eq:nbarr} radiatively generates the rolling potential. The parametric dependence illustrated in the previous section holds besides that $g_u f$ would be here bounded from below by $\mu^{d}$. Most importantly, Eq.~\eqref{upper_bound} still limits how high the UV cut-off of the Higgs sector can be.


\medskip

\subsection{Froggatt-Nielsen Charges}\label{sec:charges}

Before giving an explicit charge assignment for our setup, we review the basic requirements on the FN charges in our construction. For simplicity, we assume $g^{u,d}_{j}\sim \tilde{g}^{u,d}_{j}\sim g_{u,d}$ for all three generations, but the requirements below can be easily adapted if some flavor texture is assumed  in the $g$'s. 
\begin{enumerate}
	\item In order to get an $\mathcal{O}(1)$ CKM phase, Eq.~\eqref{eq:cond1} and Eq.~\eqref{eq:cond2} should be fulfilled at least in either the up or the down quark sector. Eq.~\eqref{eq:cond2} can be satisfied if the FN charge assignment suppresses $\mu^{u,d}$ enough. In the FN framework, Eq.~\eqref{eq:cond1} can be generalized by rescaling $g^{q}_{k}\to\varepsilon_{q}^{\left|\left[\psi_{q}\right]+\left[q^{c}_{k}\right]\right|}\cdot g^{q}_{k}$ and $\tilde{g}^{q}_{k}\to\varepsilon_{q}^{\left|\left[\psi_{q}\right]+\left[q^{c}_{k}\right]\right|}\cdot\tilde{g}^{q}_{k}\,$, where $q\in\left\{u,d\right\}\,$. The requirement in Eq.~\eqref{eq:cond1} requires than at least two generations to satisfy $\vert[\psi_{q}]+[q^{c}_{j}]\vert=\vert[\psi_{q}]+[q^{c}_{k}]\vert$ in either the up or down sector. This is achieved in the down sector in our model where all the SM quarks have the same charge. As we will see later in this subsection, this feature is completely consistent with the mixings and masses in the down sector.

	\item Eq.~\eqref{eq:ABJzero} requires a FN group with zero ABJ anomaly with QCD. This requirement immediately puts aside the standard FN scenarios, where all the charges have the same sign and the non-zero QCD anomaly can be related to the product of the determinants of up and down mass matrices~\cite{Ibanez:1994ig,Binetruy:1994ru,Binetruy:1996xk,Calibbi:2016hwq}.

	\item We always assume that the second term in Eq.~\eqref{iout} does not spoil the hierarchical structure of the first. This condition is generally true, with our assumption $g^{u,d}_{j}\sim\tilde{g}^{u,d}_{j}\sim g_{u,d}\,$, for a large class of charge assignments satisfying the following conditions ($q\in\left\{u,d\right\}$):
	\begin{equation}
	\left|[\psi_{q}]+[q^{c}_{1,2}]\right|\geq\left|[\psi_{q}]+[q^{c}_{3}]\right|\, ,\qquad  \left|[Q_{j}]+[q^{c}_{1,2}]\right|\geq\left|[Q_{j}]+[q^{c}_{3}]\right|\qquad \forall j\,.\label{eq:Quasi-Quasi-Diagonal}
	\end{equation}
\end{enumerate}

Apart form the above, the charge assignment should satisfy another requirement to keep the  higher order corrections to $\bar{\theta}_{\textrm{QCD}}$ under control. We discuss this in Appendix~\ref{app:quarks}. We are now ready to present an explicit charge assignment with $\left[u^{c}_{k}\right]\leq0\,$, that gives $n_{\text{QCD}}=0\,$, satisfies all the other requirements above,\footnote{A somewhat \emph{ad hoc} way of fulfilling Eq.~\eqref{eq:ABJzero} would be to keep a standard FN charge assignment for the SM quarks (see for example Ref.~\cite{Binetruy:1996xk}), and add new colored fermions chiral under $U(1)_m$. The contribution of the new fermions to the QCD anomaly, $\Delta n_{\text{QCD}}$, should be such that $n_{\text{QCD}}+\Delta n_{\text{QCD}}=0\,$. The typical mass of the new colored chiral fermion can be heavy enough to not pose any phenomenological challenge to this possibility.} and reproduces, with a single $5\%$ tuning, the SM quark masses and mixings. 
\begin{equation}\label{eq:quark_FN_charges1}
\left(\begin{array}{ccc}
\left[Q_1\right] & \left[Q_2\right] & \left[Q_3\right] \\
\left[u^c_1\right] &\left[u^c_2\right] &\left[u^c_3\right] \\
\left[d^c_1\right] &\left[d^c_2\right] &\left[d^c_3\right] \\
\end{array} \right) =
\left(\begin{array}{ccc}
3 & 2 & 0\\
-10 & -6& 0 \\
2 & 2 & 2 \\
\end{array} \right)\,,
\left(\begin{array}{cc}
\left[ \psi_u\right] & \left[ \psi^c_u\right] \\
\left[ \psi_d\right] & \left[ \psi^c_d\right] \\
\end{array} \right) =
\left(\begin{array}{cc}
0 & -4 \\
-2 & 6 \\
\end{array} \right)\, .
\end{equation}
The above charge assignment satisfies Eq.~\eqref{eq:Quasi-Quasi-Diagonal} above, hence the texture of the quark masses is determined by the Yukawa matrices $Y^{u,d}$ only. The parametric form for the up and down Yukawa matrices is
\begin{equation}
Y^d=\left( \begin{array}{ccc}
\varepsilon_d^{5}& \varepsilon_d^{5} & \varepsilon_d^{5}\\
\varepsilon_d^{4}&\varepsilon_d^{4} &\varepsilon_d^{4} \\
\varepsilon_d^{2} & \varepsilon_d^{2} & \varepsilon_d^{2} \\
\end{array} \right)\,,\quad
Y^u=\left( \begin{array}{ccc}
\varepsilon_u^{7}& \varepsilon_u^{3} & \varepsilon_u^{3}\\
\varepsilon_u^{8}&\varepsilon_u^{4} &\varepsilon_u^{2} \\
\varepsilon_u^{10} & \varepsilon_u^{6} & \varepsilon_u^{0} \\
\end{array} \right)\,.
\label{mats}
\end{equation}
The down Yukawa matrix, $Y^d$, has a quasi-diagonal form (\emph{i.e.},~$Y^d_{jk}/Y^d_{kk}\leq1$ for $j\leq k$), and one can read off masses and mixings parametrically:
\begin{equation}
m^d_j= Y^d_{jj}v\,,\quad \left(V^d_{L}\right)_{jk}=\frac{Y^d_{jk}}{Y^d_{kk}}\qquad\left(j\leq k\right)\,.
\label{qd}
\end{equation}
The up Yukawa matrix, $Y^u$, is instead not quasi-diagonal, unless we tune the ${\cal O}(1)$ coefficient $y^{u}_{12}$ of the 12 entry of Eq.~\eqref{mats} to be ${\cal O}(\varepsilon_u^2)$. Once we perform this tuning, we find that Eq.~\eqref{qd} holds also for the up sector with $d \to u\,$. We checked that with this $\mathcal{O}(5\%)$ tuning the charge assignment in Eq.~\eqref{eq:quark_FN_charges1} can fit all the quark masses and mixings for $\varepsilon_d=0.08$ and $\varepsilon_u=0.16$ (we used the SM Yukawas defined at the scale $10^9\textrm{ GeV}$ from Ref.~\cite{Xing:2007fb}). The above charge assignment also gives $\mu^{d}\sim \left|B^{d}_{k}\right|\,$, where we saturate the bound $g_{u,d}\sim 10^{-3}$ in ~\eq{eq:bound}. Having $\varepsilon_u\sim 2 \varepsilon_d$ translates into a mild hierarchy between the cut-off of the up and down sectors: $\Lambda_{d}\sim2\Lambda_{u}\,$. The fact that all the down quarks have the same charge follows from
\begin{equation}
\varepsilon_d^{\left|\left[d^c_j\right]-\left[d^c_k\right]\right|}=\frac{m^d_{j}}{m^d_{k} \varepsilon_d^{\left|\left[Q_j\right]-\left[Q_k\right]\right|}}\,,
\end{equation}
where the right hand side of the above equation is ${\cal O}(1)$ for all possible pairs $j<k$ for $\varepsilon_d=0.08\,$.


\medskip

\section{The Lepton Sector}\label{sec:Leptons Sector}
We can easily extend our FN construction to the charged and neutral lepton sector in a standard way, such that the Yukawa couplings in these sectors also arise from higher dimensional operators. The neutrino masses can be addressed by the seesaw mechanism, by adding sterile neutrinos that have a FN preserving Majorana mass. We refer to Ref.~\cite{Ema:2016ops} for a complete implementation of the FN mechanism in the lepton sector along these lines. To obtain the backreaction potential, we still need to introduce a $U(1)_{\text{clock}}$ breaking sector at the first site of the clockwork chain, such as the strongly coupled sector in the non-QCD model of Ref.~\cite{Graham:2015cka}, or the weakly coupled one in Ref.~\cite{Gupta:2015uea}. The backreaction scale in this scenario would depend on the details of the new sector, and can be as high as $4 \pi v$. This will thus give us a model that addresses all the SM hierarchies and neutrino masses, and thus will provide an existence proof of the hierarchion idea. 

In what follows, we pursue a conceptually more elegant and minimal alternative, where the sterile neutrino sector itself gives rise to the backreaction potential, and there is no need to introduce an extra \emph{ad hoc} sector. A unique feature of our construction is that the backreaction is generated by states which are singlet under the SM gauge group. These should be below the EW VEV $v$ (and not $4 \pi v$ like in Eq.~\eqref{eq:2loop}) to get a successful backreaction potential. The only other model where a successful relaxion mechanism is achieved with only SM singlets is the one in Ref.~\cite{Espinosa:2015eda}, which involves, however, two light axion-like states.


\medskip

\subsection{The Sterile Neutrino Backreaction}\label{sec:neutrinos}

We couple the SM leptons to the $0^\textrm{th}$ site of the clockwork construction \emph{\`a la} FN. We identify the scalar $\Phi_0$ at the $0^\textrm{th}$ site of the clockwork chain with the flavon field, which couples to SM leptons via non-renormalizable interactions, and $U_0= e^{i\pi_0/f}$ is the FN familon. Since $\langle\pi_0 |\phi\rangle\approx 1\,$, the familon of the lepton sector is identified with the relaxion and it will be approximately massless as long as the $U(1)_{\text{clock}}$ is not explicitly broken. The FN Lagrangian for the lepton sector is
\begin{align}\label{eq:fn_lep}
\mathcal{L}_{L} & = y^c_{jk}\cdot\left(\frac{\hat{\Phi}_0}{\Lambda_c}\right)^{\left|\left[L_j\right]+\left[e^c_k\right]\right|} L_{j} H e^{c}_{k} + y^n_{jk}\cdot\left(\frac{\hat{\Phi}_0}{\Lambda_{n}}\right)^{\left|\left[L_j\right]+\left[N^c_k\right]\right|}L_{j} \tilde{H} N^{c}_{k} + \textrm{h.c.}\notag\\
& \supseteq Y^{c}_{jk}\cdot U_0^{\left[L_j\right]+\left[e^c_k\right]} L_{j} H e^c_{k} + Y^{n}_{jk}\cdot U_0^{\left[L_j\right]+\left[N^c_k\right]} L_{j} \tilde{H} N^{c}_{k} + \textrm{h.c.}\,,
\end{align}
where $L$ are the $SU(2)$ doublet leptons, $e^{c}$ are the singlet leptons, and $N^{c}$ are two sterile neutrinos which are added to the SM. The charges of the fields are normalized such that $\left[\Phi_{0}\right]=-1\,$, and we define the Yukawa couplings
\begin{equation}
Y^{c}_{jk}=y^{c}_{jk}\cdot\varepsilon_{c}^{\left|\left[L_j\right]+\left[e^c_k\right]\right|}\cdot e^{i \left( \left[L_j\right]+\left[e^c_k\right] \right) \theta_0}\,,\quad Y^{n}_{jk}=y^{n}_{jk}\cdot\varepsilon_{n}^{\left|\left[L_j\right]+\left[N^c_k\right]\right|}\cdot e^{i \left( \left[L_j\right]+\left[N^c_k\right] \right) \theta_0}\,,\label{chg}
\end{equation}
in analogy with the quark sector. Terms such as $(\hat{\Phi}_0/\Lambda_n)^{\left|\left[N^c_j\right]+\left[N^c_k\right]\right|-1}\hat{\Phi}_{0}N^c_j N^c_k$ are prohibited by a lepton number symmetry, under which $L$ and $N^c$ carry opposite charges, and $\Phi_0$ is neutral.

If the sterile neutrino sector breaks the FN symmetry softly, the $U(1)_{\text{clock}}$ gets explicitly broken, and a backreaction potential is radiatively generated. As we shall discuss, the main model building challenge is to make the Higgs-dependent contributions to the potential dominating the Higgs-independent ones, such that the EW scale can be successfully selected by the relaxion mechanism. The basic idea to achieve this would be to make use of the FN charge assignment to impose a non-generic pattern in the sterile neutrino mass matrix.

We present here a simple example of how a successful backreaction sector can be achieved with only SM singlets. Two extra sterile neutrinos, $N$, are added, which have a lepton number opposite to that of the $N^c$'s. We write the following Lagrangian
\begin{align}
{\cal L}_{N}^{\text{br}} &= y^D_{jk}\cdot\left(\frac{\hat{\Phi}_0}{\Lambda_n}\right)^{\left|\left[N_j\right]+\left[N^c_k\right]\right|-1}\hat{\Phi}_0 N_{j} N_{k}^{c} + \frac{1}{2}M_{jk}^{M} N_{j} N_{k} + {\rm h.c.}\nn\\
&\supseteq M^D_{jk}\cdot U_0^{\left[N_j\right]+\left[N^c_k\right]} N_{j} N_{k}^{c} + \frac{1}{2}M_{jk}^{M} N_{j} N_{k} + {\rm h.c.}\,,
\label{fnb}
\end{align}
where we have defined the Dirac mass as $M^D_{jk}=y^{D}_{jk}\cdot\varepsilon_{n}^{\left| \left[N_j\right]+\left[N^c_k\right]\right|-1}\cdot e^{i \left(\left[N_j\right]+\left[N^c_k\right] \right) \theta_0}\frac{f}{\sqrt{2}}\,$. Notice that the Dirac mass term preserves all the symmetries, whereas the Majorana mass term for $N$ softly breaks both the lepton number and, with two or more generations, the FN symmetry.

To understand the structure of the radiatively induced potential, without loss of generality we can rotate the fields to a special, more convenient, basis, where the only non-derivative interactions of $U_0$ are controlled by the diagonal entries of the Majorana mass matrix of the two inert sterile neutrinos. In this new basis, the non-derivative familon couplings are proportional to the charge difference between the two inert sterile neutrinos $\Delta n =[N_2]-[N_1]\,$. We define the $U_0$-dependent mass matrix in this basis
\begin{equation}\label{eq:barM^M}
\overline{M}^{M}\equiv\left(\begin{array}{cc}
U_{0}^{\Delta n/2}& 0 \\
0& U_{0}^{-\Delta n/2} \\
\end{array}\right)M^{M}\left(\begin{array}{cc}
U_{0}^{\Delta n/2}& 0 \\
0& U_{0}^{-\Delta n/2} \\
\end{array}\right)\,.
\end{equation}
All the contributions to the relaxion potential are then proportional to $\cos\!\frac{\Delta n \phi}{f}$ (as expected from the fact that the breaking of the shift symmetry only arises for two or more generations). Since a collective breaking mechanism is at work, the quadratic divergent diagrams are independent of $U_0$. Integrating out the full neutrino sector at 1-loop, we get two kinds of contributions to the Coleman-Weinberg~(CW) potential: 
\begin{gather}
V_D\sim \frac{\text{Tr}(M^{D}M^{D\dagger} \overline{M}^{M} \overline{M}^{M\dagger})}{16 \pi^2}\log\frac{m^2_{\text{clock}}}{M^2}\,, \label{eq:danger}\\
V_\text{br}\sim H^{\dagger}\!H\left[\frac{\text{Tr}(Y^{n}M^{D\dagger}\overline{M}^{M}\overline{M}^{M\dagger}M^{D}Y^{n\dagger})}{16 \pi^2 M^2}+\dots\right]\,, 
\label{eq:br}
\end{gather}
where $M\approx\max\!\left[M^D,M^M\right]\,$, and we cut-off the loop integral at the scale of the clockwork radial modes, $m_{\text{clock}}\approx g_\textrm{clock} f\,$, where $g_{\textrm{clock}}$, defined in Eq.~\eqref{eq:clock_pot}, is an $\mathcal{O}(1)$ coupling. In Appendix~\ref{app:leptons}, we show how the scaling of these contributions can be easily derived in the limit $M^M\gtrsim M^D$ by integrating out the $N$'s. Another way is to directly expand the full CW potential in the mass insertion approximation. 

Notice that $V_D$ is log-enhanced, and that the $U_0$-dependent piece scales linearly with the off-diagonal entries of the Dirac mass $M^D$. Indeed, it is easy to show that if $M^D$ is diagonal, then Eq.~\eqref{eq:danger} does not depend on $U_0$.\footnote{In the special basis defined above, the matrix $\overline{M}^{M} \overline{M}^{M\dagger}$ contains $U_0$ only in the off-diagonal entries, so that $\text{Tr}[M^{D} M^{D\dagger} \overline{M}^{M} \overline{M}^{M\dagger}]$ is independent of $U_0$ as long as $M^D$ is diagonal.} $V_{\text{br}}$ is, instead, a sum of finite contributions, like the one showed in Eq.~\eqref{eq:br}, and is non-zero when the Dirac mass $M^D$ is diagonal. This difference makes it possible to parametrically suppress $V_D$ by making $M^D$ nearly diagonal, thus getting a successful backreaction potential with sterile neutrinos at the EW scale. 

For simplicity we take all the entries in $M^M$ to be of the same order, and fix $\left|\left[N_{k}\right]+\left[N^{c}_{k}\right]\right|$ to get the diagonal terms in $M^D$ of the same order:
\begin{equation}\label{eq:Soft Braking Scale Assumption}
M^M_{jk}\sim M^M_{jj}\sim M\quad ,\quad \varepsilon^{\left| \left[N_{k}\right]+\left[N^{c}_{k}\right]\right|-1}\sim M/f\,. 
\end{equation}
For ease of notation, we use the following definitions for the FN charges:
\begin{equation}
n_{jk} \equiv \left[L_j\right]+\left[N^c_k\right] \quad,\quad n_{jk}^D \equiv \left[N_j\right]+\left[N^c_k\right]\,.
\end{equation}
To get Eq.~\eqref{eq:danger} subdominant compared to Eq.~\eqref{eq:br}, we have an upper bound on the masses of the sterile neutrinos 

\begin{equation}\label{eq:Sterile Neutrino Threshold}
M\lesssim v\left(\frac{\varepsilon_{n}^{\left| n_{jj}\right|+\left| n_{jk}\right|+\left| n_{jj}^D\right|-\left| n_{jk}^D\right|}}{\log\frac{m^2_{\text{clock}}}{M^2}}\right)^{\frac{1}{2}}\approx 29\textrm{ GeV}\cdot\frac{6}{\sqrt{\log\frac{m^2_{\text{clock}}}{M^2}}}\cdot \frac{\varepsilon_{n}^{\frac{1}{2}\left(\left| n_{jj}\right|+\left| n_{jk}\right|+\left| n_{jj}^D\right|-\left| n_{jk}^D\right|\right)}}{1}\,,
\end{equation}
where $j\neq k\,$, there is no summation on repeated indices, and we have taken $m_{\textrm{clock}}\gtrsim10^{9}\text{ GeV}\,$. Notice that under these conditions, the constraint of Eq.~\eqref{eq:2loop}, coming from suppressing the 2-loop contributions to the backreaction potential, is automatically fulfilled.

The advantage of getting the backreaction potential from the sterile neutrino sector is that within the same construction, we can explain the spectrum of SM neutrino and charged leptons. Furthermore, the unique flavor structure presented above gives predictions on the absolute scale of the SM neutrino parameters. To obtain the active neutrino mass matrix, we integrate out the four sterile neutrinos, and get the Weinberg operator
\begin{equation}\label{eq:Weinberg Operator}
\mathcal{L}_\nu=-\frac{1}{2}W_{jk}\tilde{H}\tilde{H}L_j L_k\qquad\Rightarrow\qquad m_\nu\sim y_N^2 \frac{v^2}{M}\ .
\end{equation}
where $W=Y^{n}\left(M^{D}\right)^{-1}M^{\!M}(M^{D})^{-1\,T}(Y^{n})^{T}\,$, and we write on the right hand side the parametric behavior of the non-zero neutrino masses, defining $y_N$ as the order of magnitude of the elements of the Yukawa matrix $Y^n$. Since $W$ is rank 2, one of the neutrino mass eigenstates is massless. Hence, the absolute scale of the SM neutrinos mass parameters is deduced from the measurements of neutrino oscillations. The large mixing angles in the lepton sector are explained by assigning equal or similar charges to the lepton $SU(2)$ doublets (see, for instance, Refs.~\cite{Nir:2004pw,Perez:2008ee}). In the case of one massless neutrino eigenstate and inverted hierarchy structure, $\left(m_{2}-m_{1}\right)/m_{1}<2\%\,$, namely $m_{1}$ and $m_{2}$ are approximately degenerate. This approximate degeneracy cannot be accounted for by a general anarchic mass matrix, thus our model favors normal flavor ordering of neutrino masses.

The backreaction potential in Eq.~\eqref{eq:br} depends quadratically on the EW VEV. However, getting the correct SM neutrino masses provides an extra constraint on the parametric of the backreaction potential, which gives 
\begin{equation}
\Lambda_{\text{br}}\sim \left(\frac{y_N^2 v^2 M^2}{16\pi^2}\right)^{1/4}\sim \left(\frac{m_\nu M^3}{16\pi^2}\right)^{1/4}\lesssim 10\text{ MeV}\,.\label{eq:bkscaling}
\end{equation}
To get the upper bound, we used Eq.~\eqref{eq:Sterile Neutrino Threshold}, and took $m_\nu$ to be the upper bound on the sum of SM neutrino masses~\cite{Giusarma:2016phn}.

We now give a working charge assignment for the full lepton sector - one which leads to a successful backreaction, and a correct flavor structure for the SM leptons and neutrinos. We take $\varepsilon_{n}=0.09$ and $\varepsilon_{c}=0.07\,$, and assign the following $U(1)_0$ charges: 

\begin{equation}\label{eq:lepton_FN_charges}
\left(\begin{array}{ccc}
\left[ L_1\right] & \left[ L_2\right] & \left[ L_3\right] \\
\left[ e^c_1\right] &\left[ e^c_2\right] &\left[ e^c_3\right] \\
\end{array} \right) =
\left(\begin{array}{ccc}
0 & 0 & 0\\
5 & 3& 2 \\
\end{array} \right)\,,\quad
\left(\begin{array}{cc}
\left[ N_1\right] & \left[ N_2\right] \\
\left[ N^c_1\right] &\left[ N^c_2\right] \\
\end{array} \right) =
\left(\begin{array}{cc}
-15 & 15 \\
7 & -7 \\
\end{array} \right)\,.
\end{equation}
This charge assignment is valid for $M=20\textrm{ GeV}$ and $f=10^9\textrm{ GeV}\,$, as can be understood from Eq.~\eqref{eq:Soft Braking Scale Assumption}.
In this specific choice, the additional heavy neutrino mass states will have their mass values span in the range $\sim 1-100\textrm{ GeV}\,$.

Due to $\left[ L_j\right]$ being zero, the mass hierarchies of the charged leptons are realized by choosing the lepton $SU(2)$ singlets charges: $m_\mu/m_{\tau}\sim\varepsilon_c^{\left[e^c_2\right]-\left[e^c_3\right]}$ and $m_e/m_{\tau}\sim\varepsilon_c^{\left[e^c_1\right]-\left[e^c_3\right]}\,$. Given the parametric in Eq.~\eqref{eq:Weinberg Operator}, a value of $y_N\sim 10^{-7}$ gives the correct neutrino mass scale. While this is a small value for the Yukawa coupling compared to standard seesaw scenarios, it is still much larger than the size of the Yukawas for purely Dirac neutrinos, which is $y_N\sim m_{\nu}/v\sim10^{-14}\,$. The smallness of neutrino masses in our model arises then from a combination of the FN and seesaw mechanisms. Moreover, the charge assignment of the sterile neutrinos results in a nearly diagonal $M^D$ mass matrix, which is required in order to suppress the Higgs-independent terms in Eq.~\eqref{eq:danger}.

As a final remark, it is worth mentioning that since the relaxion $\phi$ gets a VEV at the end of its dynamics (see Eq.~\eqref{eq:order1}), there is a physical phase, $e^{i\theta_0}$, in the leptonic sector, which can always be rotated into diagonal entries of $M^M$. The existence of such an $\mathcal{O}(1)$ phase opens the interesting possibility that our construction could satisfy the Sakharov conditions for baryogenesis. An investigation of this idea is left for future works.


\medskip

\section{Parameter Space and Phenomenology}\label{sec:pheno}

We now discuss the phenomenological consequences of our construction, and assess its parameter space. Our results are summarized in Fig.~\ref{fig:bounds}. Like in other relaxion models, we have to deal with two constraints:
\begin{itemize}
\item There is an upper bound on $\Lambda_H$ coming from requiring a successful relaxion cosmology. As first discussed in Ref.~\cite{Graham:2015cka}, in order for the relaxion mechanism to work, two requirements must be met: first, the vacuum energy during inflation should be greater than the vacuum energy due to the relaxion field, and secondly, the time evolution of $\phi$ should be dominated by its classical rolling. Combining these requirements, one finds that
\begin{equation}
1\text{ TeV}\lesssim\Lambda_H\lesssim \left(\frac{M_{\textrm{Pl}}}{r_{\textrm{roll}}}\right)^{\frac{1}{2}}\cdot \left(\frac{\Lambda_{\textrm{br}}^4}{f}\right)^{\frac{1}{6}}\,,
\label{eq:cosmobound}
\end{equation}
where we also indicated the lower bound on $\Lambda_H$ arising from not having seen any new physics at the LHC. Notice that this upper bound can be overcome in alternative setups, where the relaxion is also the inflaton, as in Ref.~\cite{Tangarife:2017rgl}, and the first requirement is not necessary. The necessity of the second requirement is discussed in more detail in Ref.~\cite{Gupta:2018wif}.
\item There is an upper bound on $\Lambda_{\text{br}}$ which comes from the requirement of having a successful backreaction sector. The estimate from naive dimensional analysis of this bound is given by Eq.~\eqref{eq:2loop}.  In the hierarchion setup, requiring the backreaction to arise from the sterile neutrino sector implies the more stringent bound given in Eq.~\eqref{eq:bkscaling}. 
\end{itemize}
There are two extra constraints which come from the NB relaxion:
\begin{itemize}
\item The relations in Eq.~\eqref{eq:matchingH} and in Eq.~\eqref{upper_bound} provide a direct relation between the cut-off scale $\Lambda_H$ and relaxion decay constant $f$. These relations are typical of UV sensitive implementations of the NB relaxion, where the $U(1)_{\text{clock}}$ is broken \emph{explicitly} by the NB sector. 
\item In order to keep the quantum corrections of the NB construction under control, we get a strong upper bound on the couplings $g_{u,d}\,$, which we write in Eq.~\eqref{eq:bound}. The details of how this bound is obtained are in Appendix~\ref{app:quarks}. As already noticed in Ref.~\cite{Davidi:2017gir}, the bottom line is that such an upper bound is generic in any flavor construction which goes beyond Minimal Flavor Violation.

\begin{figure*}[!t]
\centering
\includegraphics[scale=0.5]{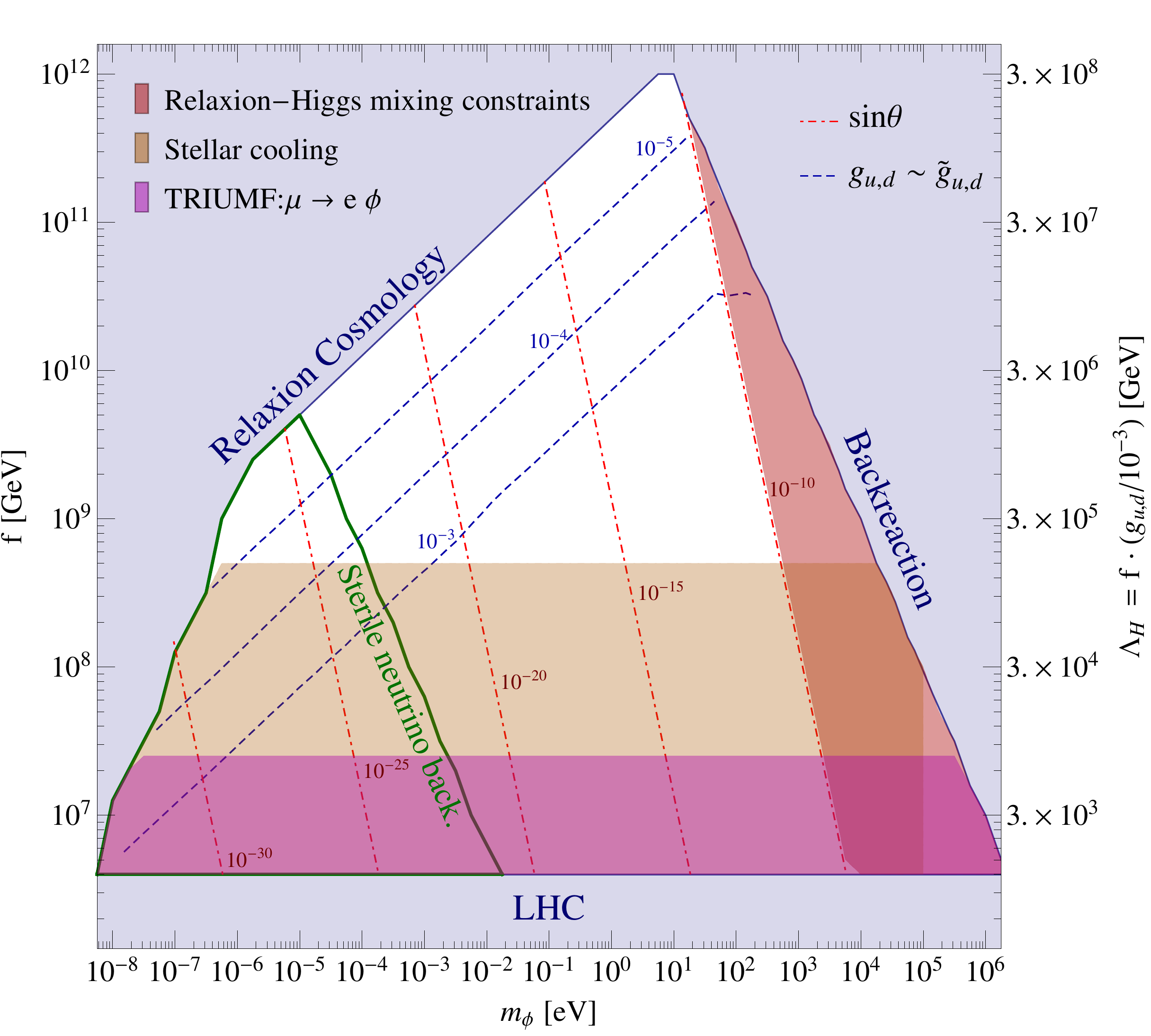}
\caption{Allowed parameter space of the hierarchion construction. We marginalize on $g^{u,d}\sim\tilde{g}^{u,d}\in(10^{-3},10^{-6})\,$. The blue dashed contours indicate the maximal value of these couplings. The blue shaded boundaries results from LHC constraints, having successful relaxion cosmology (see Eq.~\eqref{eq:cosmobound}), and the upper bound on the backreaction scale in Eq.~\eqref{eq:2loop}. In the region inside the green boundary, the backreaction is generated by the sterile neutrino portal (see Eq.~\eqref{eq:bkscaling}). We include constraints on the hierarchion from its Higgs portal couplings~\cite{Flacke:2016szy} (red), from flavor violating muon decays~\cite{Jodidio:1986mz} (magenta), and star cooling from electron coupling~\cite{Raffelt:1985nj,Raffelt:1996wa} (orange). The red dot-dashed contours correspond to the relaxion-Higgs mixing in Eq.~\eqref{eq:mixinghiggs}.}
\label{fig:bounds}
\end{figure*}
\end{itemize}
Putting all the above constraints together, we get the shaded blue triangle-shaped boundaries in Fig.~\ref{fig:bounds}. The shape of these boundaries is typical of any NB relaxion model as noticed in Ref.~\cite{Davidi:2017gir}. The boundary on the right is given by the upper bound on the backreaction scale. The smaller green triangle-shaped region in Fig.~\ref{fig:bounds} corresponds to the region of the parameter space where the sterile neutrino portal can generate the backreaction scale, and Eq.~\eqref{eq:bkscaling} is satisfied. The left and bottom boundaries of the triangle are given by the cut-off constraints in Eq.~\eqref{eq:cosmobound}. Through the matching conditions in Eq.~\eqref{eq:matchingH} and in Eq.~\eqref{upper_bound}, the left boundary is dependent on the value of $g^{u,d}\sim\tilde{g}^{u,d}\sim g_{u,d}\lesssim 10^{-3}\,$. In Fig.~\ref{fig:bounds} we marginalize over these couplings, indicating in the blue contours their maximal value in every region of the parameter space. The Higgs UV cut-off is strongly bounded from above for the sterile neutrino backreaction sector, but can be much higher if a new backreaction sector is introduced
\begin{equation}
\!\!\!\!\Lambda_H\lesssim (10^2,\,10^5)\textrm{ TeV}
\!\cdot\!\left(\frac{g^u\tilde{g}^u}{10^{-6}}\right)^{\!\!\frac{2}{7}}
\!\!\cdot\!\left(\frac{(Y^{u})^{T}Y^u}{1}\right)^{\!\!\frac{1}{2}}
\!\!\cdot\!\left(\frac{\log\frac{g_{\textrm{clock}}}{g^{u}}}{5}\right)^{\!\!\frac{1}{2}}
\!\!\cdot\!\left(\frac{\Lambda_{\textrm{br}}}{10\textrm{ MeV},\,100\textrm{ GeV}}\right)^{\!\!\frac{4}{7}}\,.\!\!\!\!\label{eq:cutoffbound1}
\end{equation}
Two related drawbacks of the hierarchion setup are the increased number of clockwork sites compared to other relaxion scenarios, and the theoretical challenge of screening the relaxion backreaction potential from Planck scale suppressed corrections. We refer to Ref.~\cite{Davidi:2017gir} for a discussion of this last point.

We now comment on the phenomenology of our setup. The mass of the hierarchion is set by the backreaction sector, so that we get
\begin{equation}
10^{-8}\lesssim m_\phi\simeq \frac{\Lambda_{\text{br}}^2}{f}\lesssim 10^6\, ,10^{-2} \text{ eV}\,,
\end{equation}
where the two values on the right hand side of the inequality correspond to the highest possible backreaction scale and to the sterile neutrino backreaction, respectively. Within this mass range, the hierarchion is always long lived, so it will show up as missing energy at collider experiments, and possibly affect astrophysical and cosmological processes. The main novelty of the hierarchion is that it carries the standard Higgs portal couplings of any relaxion model together with the familon couplings typical of FN constructions based on global symmetries~\cite{Feng:1997tn,Calibbi:2016hwq,Ema:2016ops}. We briefly describe the features of these two set of couplings in turn: 
\begin{itemize}
\item The Higgs portal coupling are generated from the backreaction sector, as described in Ref.~\cite{Flacke:2016szy}. The relaxion mixing with the SM Higgs can be written as 
\begin{equation}
\sin\theta\simeq \frac{\Lambda_{\text{br}}^4}{f v m_h^2}\ ,\label{eq:mixinghiggs}
\end{equation}
and it is shown in Fig.~\ref{fig:bounds} as dot-dashed red contours. In our particular mass range, the relevant constraints exclude the red-shaded region in Fig.~\ref{fig:bounds}. This is the rough combination of bounds coming from astrophysical processes~\cite{Raffelt:1987yt,Raffelt:2012sp,Hardy:2016kme,Chang:2018rso}, from distortion of the Extragalactic Background Light~\cite{Flacke:2016szy}, and from flavor-violating Kaon decays induced by the Higgs portal couplings~\cite{Adler:2008zza}.
\item The familon-type couplings result in derivative interactions of the hierarchion with SM fermions, which induce $\Delta F=1$ FV processes at tree-level. It is convenient to summarize here the couplings of the hierarchion to SM fermions after EWSB:
\begin{equation}
\!\!\!\!\mathcal{L}_\phi\!\supset\!\frac{iv}{f}\phi\!\left[\left(\left[L_j\right]\!+\!\left[e^c_k\right]\right)\!\cdot\!Y^{c}_{jk}e_{j}e^{c}_{k}\!+\!\frac{\left[Q_{j}\right]\!+\!\left[u^{c}_{k}\right]}{3^{m}}\!\cdot\!Y^{u}_{jk}u_{j}u^{c}_{k}\!+\!\frac{\left[Q_{j}\right]\!+\!\left[d^{c}_{k}\right]}{3^{m}}\!\cdot\!Y^{d}_{jk}d_{j}d^{c}_{k}\right]\!+\!\text{h.c.}\,,\!\!\!\!\label{eq:couplings}
\end{equation}
In the mass eigenbasis, the $Y^{f}_{jk}$ ($f=c,u,d$) are diagonal by definition, however the familon couplings are not unless the charge matrix is proportional to the identity. Notice that if we want to generate the backreaction through the sterile neutrino sector, then the lepton sector needs to be localized at the $0^{\textrm{th}}$ site, and the lepton FN charges are unsuppressed. We can estimate the width of the FV lepton decays as
\begin{align}
&\Gamma(\mu\to e\,\phi)\approx\frac{m_e^2m_\mu}{16\pi f^2}\,,\label{eq:muon}\\
&\Gamma(\tau\to \mu\,\phi)\approx\Gamma(\tau\to e\,\phi)\approx\frac{m_\mu^2m_\tau}{16\pi f^2}\,.\label{eq:tau}
\end{align}
Notice that the FV decay widths of the muon and the tau are suppressed with respect to the generic estimates of Ref.~\cite{Feng:1997tn}, by $\sim(m_e/m_\mu)^2$ and $\sim(m_\mu/m_\tau)^2\,$, respectively. This is because when $\vec{\left[L\right]}=(0,0,0)\,$, we need to pay a right-handed rotation to get a FV coupling. As the right-handed charges need to reproduce the lepton mass hierarchy, the 12 and 23 off-diagonal terms in Eq.~\eqref{eq:couplings} are suppressed by $\sim m_e /m_\mu$ and $m_\mu/m_\tau\,$, respectively.\footnote{The same conclusion holds for $\vec{\left[L\right]}=(q_L,q_L,q_L)$ with $q_L\neq0\,$, where the FV decays come from the right-handed rotation only.} In Fig.~\ref{fig:bounds}, we show the current best bound on $\text{BR}(\mu^{+}\!\to\! e^{+}\,\phi)< 2.6\cdot10^{-6}$ from the TRIUMF experiment~\cite{Jodidio:1986mz}, which translates into $f\gtrsim 2.8\cdot 10^7\text{ GeV}\,$. Bounds from FV tau decay gives subdominant constraints. We comment in Appendix~\ref{app:lepton_pheno} on possible improvements in searches of FV decays in the lepton sector at Belle~II, MEG~II, and Mu3e.
\item The flavor diagonal couplings of the hierarchion to SM electrons affect star cooling, resulting in $f\gtrsim 6\cdot 10^8\text{ GeV}$~\cite{Raffelt:1985nj,Raffelt:1996wa} up to relaxion masses of 100 keV. This constraint is almost two orders of magnitude stronger than the one from FV leptonic decays when $\vec{\left[L\right]}=(0,0,0)\,$. This constraint severely reduces the parameter space for the sterile neutrino portal.
\item The couplings to SM charged quarks are suppressed by $1/3^m$ compared to the ones of the SM leptons, because of the small overlap between $\phi$ and $\pi_m$. Generically, the dominant constraint on FV decays is the one of the charged Kaon 
\begin{equation}
\Gamma(K^{+}\to\pi^{+}\,\phi)\approx \frac{m_K}{64\pi} B_s^2\left[1-\frac{m_\pi^2}{m_K^2}\right]\frac{m_s m_d}{ f_m^2}\,,
\end{equation}
where $f_m=3^{-m}f\,$, and $B_s=4.6$ is the non-perturbative parameter related to the quark condensate~\cite{Kamenik:2011vy}. Combined E787 and E949 data~\cite{Adler:2008zza} give $\text{BR}(K^{+}\to\pi^{+}\,\phi)<7.3\cdot 10^{-11}\,$, which implies $f_m\gtrsim 8\cdot 10^{10}$ GeV. This becomes the strongest bound on $f$ as long as $m\lesssim 2-3\,$. Current Kaon experiments, like NA62 for $K^{+}$~\cite{Fantechi:2014hqa} and KOTO for $K_L$~\cite{Tung:2016xtx}, will extend the reach on $f_m$ of about two orders of magnitude. Complementary bounds can be obtained from FV decays of B-mesons, which, however, are sensibly weaker than the ones from Kaons, and do not pose any further constraint on our construction.
\end{itemize}

The finite allowed region in Fig.~\ref{fig:bounds} is a notable feature of our construction, implying that future experiments can potentially probe the full parameter space of the hierarchion. This is especially true for the sterile neutrino backreaction model (the dark green triangle in Fig.~\ref{fig:bounds}), which could be possibly discovered/excluded by the next generation of experiments probing lepton flavor violation, if dedicated triggers on electron-only events will be developed. In order to get enough background rejection on electron-only events, an upgrade of the data-acquisition system might be needed (see for example Ref.~\cite{talk} for a discussion on Mu3e). We further comment on these issues in Appendix~\ref{app:lepton_pheno}, and defer a more quantitative study  for a future work.   


\medskip

\section{Conclusions}

In this paper, we presented a relaxion model where the quark flavor textures and the smallness of the strong CP phase are also accounted for. We also discussed two ways of embedding the charged lepton flavor textures and neutrino masses in our construction. We showed how sterile neutrinos can generate the ``backreaction'' potential, tying together the relaxion mass and the Standard Model neutrino masses. The sterile neutrino backreaction is a unique example of how the relaxion potential can be generated without adding new electroweak~(EW) charged states around the weak scale. 

Models addressing all the Standard Model problems together already appeared in the context of standard solutions of the hierarchy problem, like Supersymmetry or Compositeness. However, in these setups there is always a certain amount of tension between CP and flavor observables, and the naturalness of the EW scale, which is often dubbed as ``the new physics CP and flavor puzzles''~\cite{Nir:2007xn}. A key difference of the hierarchion construction is that flavor and CP violating processes are not an issue since there is no cost in pushing up the scale of new physics if the relaxion mechanism is at work. 

Following the Nelson-Barr~(NB) relaxion idea~\cite{Davidi:2017gir}, the $\mathcal{O}(1)$ phase of the relaxion, which was causing the ``relaxion CP problem'', becomes a natural source for the Cabibbo-Kobayashi-Maskawa~(CKM) phase in our model. The large hierarchy between the small $\bar{\theta}_{\rm QCD}$ and the large CKM phase is guaranteed by the NB mechanism. In this paper, we further showed how the NB relaxion can be embedded in a full Froggatt-Nielsen construction, explaining the quark masses and flavor textures at the price of a single small coupling controlling the relaxion ``rolling'' potential $g^{u,d}\sim \tilde{g}^{u,d}\lesssim 10^{-3}$. In the explicit charge assignment we showed, a residual 5\% tuning in the up quark matrix is left unexplained. This issue could be possibly solved in more general charge assignments. 

Our construction gives an upper bound on the scale of colored states of 10-1000 TeV, depending on the hierarchion decay constant. This bound motivates future colliders, directly testing energy scales beyond the reach of the LHC. The future reach of collider searches has also an interesting interplay with the searches for new light states with flavor violating~(FV) couplings. Indeed, the hierarchion is generically a light familon, and gives a strong motivation to extend the coverage and reach of precision measurements looking for rare FV decays of leptons and mesons. In particular, upcoming experiments, like MEG~II and Mu3e, might have the possibility of probing the parameter space of the hierarchion beyond the present best constraints, given by star cooling. A sufficient increase in sensitivity of these experiments in electron-only events could possibly probe the full parameter space of the hierarchion in its most interesting realization.


\medskip

\subsection*{Acknowledgments}

We thank Lorenzo Calibbi, Yossi Nir, and Lorenzo Ubaldi for useful discussions. 
The work of GP is supported by grants from the BSF, ERC, ISF, Minerva, and the Weizmann-UK Making Connections Program.


\medskip

\appendix

\section{Higher Order Contributions to $\bar{\theta}_{\textrm{QCD}}$}\label{app:quarks}

As shown in Eq.~\eqref{eq:arg}, the structure of the mass matrices in Eq.~\eqref{eq:quark_mass_tree_level} leads to vanishing tree-level contribution to $\bar{\theta}_\textrm{QCD}$. However, higher order effects can spoil the delicate structure of the mass matrices, giving rise to non-zero $\bar{\theta}_\textrm{QCD}$. The quantum corrections to the quarks mass matrix can be parametrized as
\begin{equation}\label{eq:mass_matrix}
M^{q}+\Delta M^{q}\equiv M^{q}+\left(\begin{array}{cc}
\left(\delta\mu^{q}\right)_{1\times1} & \left(\delta B^{q}\right)_{1\times3}\\
\left( v\delta Y^{\psi_{q}^{c}}\right)_{3\times 1} & \left(v\delta Y^{q}\right)_{3\times3}
\end{array}\right)\,,
\end{equation}
where here, and throughout this entire section, $q\in\left\{u,d\right\}\,$. Assuming $\Delta M^{q} \ll M^{q}\,$, the contribution to $\bar{\theta}_\textrm{QCD}$ reads
\begin{equation}\label{eq:Tr_identity}
{\rm Arg}\left[\det\left(M^{q}+\Delta M^{q}\right)\right]\approx\Im\left\{\Tr\left[\left(M^{q}\right)^{-1}\Delta M^{q}\right]\right\}\,,
\end{equation}
where
\begin{equation}\label{eq:M^q Inverse}
\left(M^{q}\right)^{-1}=\left(\begin{array}{cc}
\left(\mu^{q}\right)^{-1}_{1\times1} & \left(-\left(\mu^{q}\right)^{-1}B^{q}\left(vY^{q}\right)^{-1}\right)_{1\times3}\\
\left(0\right)_{3\times1} & \left(vY^{q} \right)^{-1}_{3\times3}
\end{array}\right)\,.
\end{equation}
Inserting Eq.~\eqref{eq:quark_mass_tree_level} and Eq.~\eqref{eq:M^q Inverse} into Eq.~\eqref{eq:Tr_identity}, we get three types of dangerous contributions to the strong CP phase
\begin{align}
\delta\bar{\theta}^{q}_{1} & = \Im\left\{\textrm{Tr}\left[\left(\mu^{q}\right)^{-1}\delta\mu^{q}\right]\right\}\,,\label{eq:delta theta 1}\\
\delta\bar{\theta}^{q}_{2} & = -\Im\left\{\textrm{Tr}\left[\left(\mu^{q}\right)^{-1}B^{q}\left(Y^{q}\right)^{-1}\delta Y^{\psi_{q}^{c}}\right]\right\}\,,\label{eq:delta theta 2}\\
\delta\bar{\theta}^{q}_{3} & = \Im\left\{\textrm{Tr}\left[\left(Y^{q}\right)^{-1}\delta Y^{q}\right]\right\}\,.\label{eq:delta theta 3}
\end{align}
These contributions can be caused by higher order effects arising from integrating out the FN states at the cut-off scales $\Lambda_{u,d}\sim f/\varepsilon_{u,d}\,$, or the heavy CP-even and CP-odd clockwork scalars at the scales $g_{\textrm{clock}} f$ and $\sqrt{\epsilon} f$ respectively.


\medskip

\paragraph{Threshold Effects at the Froggatt-Nielsen Scale}

We first consider the higher order effects arising from integrating out the FN states at the scales $\Lambda_{u,d}$. We assume that in the full UV theory, above the scales $\Lambda_{u,d}\,$, there are couplings that break the $U(1)_{N}$, and generate $g_{j}^{q}$ and $\tilde{g}_{j}^{q}$, the only explicit breaking couplings in the IR. We denote the UV explicit breaking couplings by $\gamma^{u,d}_{\alpha}$ and $\tilde{\gamma}^{u,d}_{\alpha}$, and assume they obey the same selection rules as $g^{u,d}_{j}$ and $\tilde{g}^{u,d}_{j}$. We can thus write the $g$'s, to leading order, as a linear combination of the $\gamma$'s
\begin{equation}\label{eq:g UV Expansion}
g^{u,d}_{j}=a^{u,d}_{j\alpha}\gamma^{u,d}_{\alpha}\,,\quad \tilde{g}^{u,d}_{j}=\tilde{a}^{u,d}_{j\alpha}\tilde{\gamma}^{u,d}_{\alpha}\,,
\end{equation}
where the elements of $a^{u,d}$ and $\tilde{a}^{u,d}$ are of order unity (see Sec.~\ref{app:UV} for a simple example of a UV completion). Keeping  this caveat in mind, from now on, for simplicity, we will use the IR $g$'s  instead of the UV $\gamma$'s. 

\begin{table}[tb]
	\centering
	\begin{tabular}{c c c}
		\hline
		& $U\left(1\right)_{\textrm{clock}}$	& $U\left(1\right)_{\psi}$	\\ [0.5ex]
		\hline
		$\Phi_{k},\,U_{k}$	& $-3^{-k}$								& $0$						\\
		$\psi_{q}$			& $\left[\psi_{q}\right]$				& $1$						\\
		$\psi_{q}^{c}$		& $\left[\psi_{q}^{c}\right]$			& $-1$						\\
		$g^{q}_{j}$			& $3^{-N}$								& $-1$						\\
		$\tilde{g}^{q}_{j}$	& $-3^{-N}$								& $-1$						\\ [1ex]
		\hline
	\end{tabular}
	\caption{Charge assignment for spurion analysis. Notice that the brackets denote the charge under the $U\left(1\right)_{\textrm{clock}}$, and not the charges under the $U\left(1\right)_{m}$, which were presented in Eq.~\eqref{eq:quark_FN_charges1}.}
	\label{table:spurion}
\end{table}
The transformation laws of Table~\ref{table:spurion} completely determine the parametric form of the higher order corrections.  At leading order in $g^{q}_{j}\sim\tilde{g}^{q}_{j}\sim g_{q}$, we get
\begin{align}
\delta\mu^{q}&\sim\left|\left(\frac{\langle\hat{\Phi}_{m}\rangle}{\Lambda_{q}}\right)^{\left|\left[\psi_{q}\right]+\left[\psi_{q}^{c}\right]\right|-1}\langle\hat{\Phi}_{m}\rangle\right| b^{q}_{jk}\left[g^{q}_{j}\tilde{g}^{q*}_{k} \left(\frac{\langle\Phi_{N}\rangle}{\Lambda_{q}}\right)^{2}+{\cal P}\right]\,,\label{eq:mur}\\
\delta Y^{\psi_{q}^{c}}_{j}&\sim\left|\left(\frac{\langle\hat{\Phi}_{m}\rangle}{\Lambda_{q}}\right)^{\left|\left[Q_{j}\right]+\left[\psi_{q}^{c}\right]\right|}\right|\left[\sum_{k} \tilde{g}^{q*}_{k}\frac{\langle\Phi_{N}\rangle}{\Lambda_{q}}+{\cal P}\right]\,,\label{eq:dyp}\\
\delta Y^{q}_{ij}&\sim\left|\left(\frac{\langle\hat{\Phi}_{m}\rangle}{\Lambda_{q}}\right)^{\left|\left[Q_{i}\right]+\left[q^{c}_{k}\right]\right|}\right|\left[g^{q}_{j}\tilde{g}^{q*}_{k}\left(\frac{\langle\Phi_{N}\rangle}{\Lambda_{q}}\right)^{2}+{\cal P}\right]\,.\label{eq:dyd}
\end{align}
In the above, $\hat{\Phi}_{m}^{\left|x\right|}=\Theta\left(x\right)\Phi_{m}^{\left|x\right|}+\Theta\left(-x\right)\left(\Phi_{m}^{*}\right)^{\left|x\right|}\,$, and ${\cal P}$ is a shorthand for an interchanging operation $\left(g^{q}_{j}, \Phi_{N}, U_{N}\right)\leftrightarrow\left(\tilde{g}^{q}_{j}, \Phi_{N}^{*},U_{N}^{*}\right)\,$. Furthermore, only powers of the modulus $\left|\langle\Phi_{m}\rangle\right|$ appear in the expressions above, since the quark sector respects the $U(1)_{m}$ symmetry, and has a zero anomaly coefficient, so the complex phase can always be rotated away. Although all couplings are real in our model, and the phase of $\Phi_{m}$ can be rotated away, maintaining the distinction between $g^{q}_{j}/\tilde{g}^{q}_{j}/\Phi_{m}$ and $g^{q*}_{j}/\tilde{g}^{q*}_{j}/\Phi_{m}^{*}$ above is useful since they carry opposite charges under the transformations of Table~\ref{table:spurion}.

The most dangerous contributions come from $\delta \bar{\theta}^{q}_{2}$ and $\delta \bar{\theta}^{q}_{3}$, where inverse powers of $\varepsilon_{q}$ arising from $\left(Y^{q}\right)^{-1}$ can potentially lead to dangerous FN enhancements of the strong CP phase. The FN quark charges thus control not only the flavor structure but also the magnitude of $\bar{\theta}_\textrm{QCD}$. We now show that the charges in Eq.~\eqref{eq:quark_FN_charges1} have been chosen such that these enhancements do not occur. In order to estimate $\delta\bar{\theta}^{q}_{2}$ and $\delta \bar{\theta}^{q}_{3}$, we use
\begin{equation}
\left(Y^{q}\right)^{-1} = \frac{\textrm{Adjugate}\left(Y^{q}\right)}{\textrm{det}\left(Y^{q}\right)}\,.
\end{equation}
For $\delta \bar{\theta}^{q}_{3}$,  the FN enhancement is proportional to $\Tr\left[\left(Y^{q}\right)^{-1}_{ik}\varepsilon_{q}^{\left|\left[Q_{k}\right]+\left[q^{c}_{j}\right]\right|}\right]\,$, which is ${\cal O}(1)$ for the charge assignment in Eq.~\eqref{eq:quark_FN_charges1}. Similarly, for $\delta\bar{\theta}^{q}_{2}$ we have
\begin{equation}
\Tr\left[\left(\mu^{q}\right)^{-1}B^{q}\left(Y^{q}\right)^{-1}\delta Y^{\psi_{q}^{c}}\right]\sim\beta^{q}_{jk}\left[g^{q}_{j}\tilde{g}^{q}_{k}\left(\frac{\langle\Phi_{N}\rangle}{\Lambda_{q}}\right)^{2}+{\cal P}\right]
\end{equation}
for the charge assignment in Eq.~\eqref{eq:quark_FN_charges1}, where $\beta^{q}_{jk}$ are ${\cal O}(1)$ numbers. Regarding $\delta \bar{\theta}^{q}_{1}$, the contribution to the imaginary part of Eq.~\eqref{eq:mur} vanishes if $b_{jk}^{q}\propto \delta_{jk}$. On pure symmetry grounds, $b_{jk}^{q}$ is a generic matrix with $\mathcal{O}(1)$ entries, so the contributions from $\delta\bar{\theta}^{q}_{1}$ are of the same order as the ones from $\delta \bar{\theta}^{q}_{2,3}$. However, for the simple UV completion presented in the following appendix, $b_{jk}^{q}-\delta_{jk}$ is further suppressed by a loop factor or by $\varepsilon_{q}^{2}$. 

Finally, note that any diagram that generates the dangerous operators above must involve the couplings $g^{q}_{j}$ and $\tilde{g}^{q}_{j}$, and ultimately light $\psi_{q}$'s or $q^{c}$'s propagators. Local contributions will thus be generated at least at 1-loop level. The dominant contributions to $\bar{\theta}_\textrm{QCD}$ scale as
\begin{equation}\label{eq:finalcorr}
\Delta\bar{\theta}_\textrm{QCD}^{\textrm{FN}}\sim\sum_{q}\frac{\alpha^{q}_{jk}}{16 \pi^2}\left(g^{q}_{j}\tilde{g}^{q}_{k}-g^{q}_{k}\tilde{g}^{q}_{j}\right)\frac{\left|\langle\Phi_{N}\rangle\right|^{2}}{\Lambda_{q}^{2}}\sin\left(2\theta_{N}\right)\,,
\end{equation}
where $\alpha^{q}_{jk}$ are, again, ${\cal O}(1)$ numbers. The experimental bound, $\bar{\theta}_{\textrm{QCD}}<10^{-10}\,$, translates into the constraint
\begin{equation}\label{eq:finbound}
g^{q}\sim \tilde{g}^q \lesssim 10^{-3} \cdot \left(\frac{0.1}{\varepsilon_{q}}\right)\,.
\end{equation}
We then assume the couplings that give rise to $g^q$ in the UV are small enough to satisfy this bound. By assuming this, we are able to address all the SM Yukawa hierarchies together with the strong CP problem.


\medskip

\paragraph{Threshold Effects at the Clockwork Scale}

We now estimate the threshold contributions to $\bar{\theta}_\textrm{QCD}$, that arise when the heavy radial and angular modes of the clockwork chain are integrated out. Barring the direct coupling of the radial and angular modes of $\Phi_{m}$ to $Q_{j} q^{c}_{k}$, the Lagrangian respects an extended version of minimal flavor violation~(EMFV), \emph{i.e.},~any flavor violation arises from the SM Yukawa matrices $Y^{q}$, $\vec{g}^{\,q}$, or $\vec{\tilde{g}}^{\,q}$ (see standard discussion in Refs.~\cite{DAmbrosio:2002vsn,Kagan:2009bn}). The non-EMFV contributions arise from the $\rho_{m} Q_{j} q^{c}_{k}$ and $\pi_{m} Q_{j} q^{c}_{k}$ couplings, which are proportional to $\left(\left[Q_{j}\right]+\left[q^{c}_{k}\right]\right)\cdot Y^{q}_{jk}\,$. We checked that the contributions involving these couplings are subdominant compared to EMFV contributions.

In the low energy theory below the scale of the clockwork pNGB's, the EMFV contributions are again determined parametrically by the selection rules in Table~\ref{table:spurion}
\begin{align}
\delta\mu^{q}&\propto\mu^{q}\left[g^{q}_{j} \tilde{g}^{q*}_{j} \langle U_{N}\rangle^{2}+{\cal P}\right]\,,\label{eq:mur2}\\
\delta Y^{\psi_{q}^{c}}_{j} &\propto Y^{q}_{jk}\frac{\mu^{q}}{f}\left[\tilde{g}^{q*}_{k} \langle U_{N}\rangle+{\cal P}\right]\,,\label{eq:dyp2}\\
\delta Y^{q}_{ij}&\propto Y^{q}_{ik}\left[g^{q}_{j}\tilde{g}^{q*}_{k} \langle U_{N}\rangle^{2}+{\cal P}\right]\,,\label{eq:dyd2}
\end{align}
where we have also used the fact that any contribution to a coupling involving $\psi_{q}^{c}$ must involve $\mu^{q}$. Notice that EMFV implies the appearance of the Yukawa matrices, $Y^{q}$, in Eq.~\eqref{eq:dyp2} and Eq.~\eqref{eq:dyd2} instead of general powers of $\varepsilon_{q}$ in Eq.~\eqref{eq:dyp} and Eq.~\eqref{eq:dyd}. This important distinction from the previous case results in the vanishing of all $\delta\bar{\theta}^{q}_{j}$ in Eqs.~(\ref{eq:delta theta 1}-\ref{eq:delta theta 3}), and we find no contribution to $\bar{\theta}_{\textrm{QCD}}$ at ${\cal O}(g_{q}^{2})$
\begin{equation}
\Delta\bar{\theta}_{\textrm{QCD}}^{\textrm{clock}}\sim\sum_{j,q}\delta\bar{\theta}^{q}_{j}\sim\Im\left[g^{q}_{j}\tilde{g}^{q*}_{j}\langle U_{N}\rangle^{2}+g^{q*}_{j}\tilde{g}^{q}_{j}\langle U_{N}^{*}\rangle^{2}\right]=0\,.
\end{equation}
The leading EMFV contribution is then ${\cal O}(g_{q}^4)$, and 2-loop suppressed~\cite{Bento:1991ez,Dine:2015jga}.\footnote{The 1-loop contributions mentioned in these references actually arise at 2-loop level in our model - this is because these diagrams include $U\left(1\right)_{\rm clock}$ breaking quartic couplings involving $\Phi_{N}$ and $H$, which are themselves generated only at 1-loop level.}


\medskip

\section{Renormalizable UV Completion of the Quark Sector}\label{app:UV}

Let us present a simple example of a renormalizable UV model, focusing on the non-trivial quark sector. As was stated in Sec.~\ref{app:quarks}, the $U(1)_{N}$ should be broken in the UV by couplings that obey the selection rules in Table~\ref{table:spurion}. After integration out of the heavy FN intermediate fields, the effective operators of Eq.~\eqref{eq:fn_quark} and Eq.~\eqref{eq:nbarr} must be recovered, where Eq.~\eqref{eq:nbarr} is the only source of $U(1)_{N}$ breaking in the IR. In fact, the most standard way of UV completing these FN operators already provides such an example. We introduce eight different kinds of heavy vector-like quark chains, all are localized at the $m^{\textrm{th}}$ site of the clockwork construction: $\left\lbrace\bar{D}_{A}\right\rbrace$, $\left\lbrace\bar{U}_{A}\right\rbrace$, $\left\lbrace\bar{D}^{c}_{A}\right\rbrace$, $\left\lbrace\bar{U}^{c}_{A}\right\rbrace$, $\left\lbrace\tilde{D}_{A}\right\rbrace$, $\left\lbrace\tilde{U}_{A}\right\rbrace$, $\left\lbrace\tilde{D}^{c}_{A}\right\rbrace$, and $\left\lbrace\tilde{U}^{c}_{A}\right\rbrace$, where $A$ is a generation label. The heavy quarks $D^{c}/U^{c}$ are in the same gauge group representation as the SM down/up quarks, $d^{c}/u^{c}$, and the heavy quarks $D/U$ are in the conjugate gauge representation. The bar denotes fields which are even under the imposed $\mathbbm{Z}_{2}$ symmetry, while the tilde denotes fields which are odd. By convention, the $U(1)_{m}$ FN charges of the new fields are represented by their subscript (\emph{e.g.},~the pair $\bar{D}_{A}\bar{D}^{c}_{-A}$ transforms as a scalar). Finally, in order to ensure that contributions to $\bar{\theta}_{\textrm{QCD}}$ arise only at loop-level, we assume that fields with the same quantum numbers as $\psi_{u,d}$ do not exist. Given this field content, the most general renormalizable Lagrangian that respects all the symmetries, is
\begin{align}
\nonumber\mathcal{L}_{\textrm{UV}}=&\;M^{\bar{D}}_{A}\bar{D}_{A}\bar{D}^{c}_{-A}+M^{\bar{U}}_{A}\bar{U}_{A}\bar{U}^{c}_{-A}+M^{\tilde{D}}_{A}\tilde{D}_{A}\tilde{D}^{c}_{-A}+M^{\tilde{U}}_{A}\tilde{U}_{A}\tilde{U}^{c}_{-A}\\
\nonumber&+\left[y^{Qd}_{j}Q_{j}H\bar{D}^{c}_{-[Q_{j}]}+y^{Qu}_{j}Q_{j}\tilde{H}\bar{U}^{c}_{-[Q_{j}]}\right]\\
\nonumber&+\left[y^{d^{c}+}_{j}\Phi_{m}\bar{D}_{-[d^{c}_{j}]+1}d^{c}_{j}+y^{d^{c}-}_{j}\Phi_{m}^{*}\bar{D}_{-[d^{c}_{j}]-1}d^{c}_{j}+\left(\bar{D},d^{c}\to\bar{U},u^{c}\right)\right]\\
\nonumber&+\left[y^{\psi_{d}+}\Phi_{m}\psi_{d}\tilde{D}^{c}_{-[\psi_{d}]+1}+y^{\psi_{d}-}\Phi_{m}^{*}\psi_{d}\tilde{D}^{c}_{-[\psi_{d}]-1}+\left(\psi_{d},\tilde{D}^{c}\to\psi_{u},\tilde{U}^{c}\right)\right]\\
\nonumber&+\left[y^{\psi_{d}^{c}+}\Phi_{m}\tilde{D}_{-[\psi_{d}^{c}]+1}\psi_{d}^{c}+y^{\psi_{d}^{c}-}\Phi_{m}^{*}\tilde{D}_{-[\psi_{d}^{c}]-1}\psi_{d}^{c}+\left(\tilde{D},\psi_{d}^{c}\to\tilde{U},\psi_{u}^{c}\right)\right]\\
\nonumber&+\left[y^{\bar{D}+}_{A}\Phi_{m}\bar{D}_{A}\bar{D}^{c}_{-A+1}+y^{\bar{D}-}_{A}\Phi_{m}^{*}\bar{D}_{A}\bar{D}^{c}_{-A-1}+\left(\bar{D},\bar{D}^{c}\to\bar{U},\bar{U}^{c}\right)\right]\\
&+\left[y^{\tilde{D}+}_{A}\Phi_{m}\tilde{D}_{A}\tilde{D}^{c}_{-A+1}+y^{\tilde{D}-}_{A}\Phi_{m}^{*}\tilde{D}_{A}\tilde{D}^{c}_{-A-1}+\left(\tilde{D},\tilde{D}^{c}\to\tilde{U},\tilde{U}^{c}\right)\right]+\textrm{h.c.}\,,
\end{align}
where $M^{\bar{D}}_{A}\sim M^{\tilde{D}}_{A}\sim\Lambda_{d}$ and $M^{\bar{U}}_{A}\sim M^{\tilde{U}}_{A}\sim\Lambda_{u}\,$. Note that the Lagrangian is written in the mass basis, with the massless quarks being identified as $d^{c}$, $u^{c}$, $\psi_{u,d}$ and $\psi_{u,d}^{c}$. Additional terms in the UV Lagrangian are those that explicitly break the $U(1)_{N}$ symmetry
\begin{align}
\nonumber\mathcal{L}_{\textrm{UV}}^{\textrm{roll}}=&\;\left(\gamma^{d}_{j}\Phi_{N}+\tilde{\gamma}^{d}_{j}\Phi_{N}^{*}\right)\psi_{d}d^{c}_{j}+\left(\gamma^{u}_{3}\Phi_{N}+\tilde{\gamma}^{u}_{3}\Phi_{N}^{*}\right)\psi_{u}u^{c}_{3}\\
&+\left(\gamma^{d}_{4}\Phi_{N}+\tilde{\gamma}^{d}_{4}\Phi_{N}^{*}\right)\psi_{d}\bar{D}^{c}_{[d^{c}_{1}]}+\left(\gamma^{u}_{4}\Phi_{N}+\tilde{\gamma}^{u}_{4}\Phi_{N}^{*}\right)\psi_{u}\bar{U}^{c}_{[u^{c}_{3}]}+\textrm{h.c.}\,,
\end{align}
where we have assumed for simplicity $\left[\psi_{u}\right]=-\left[u^{c}_{3}\right]\neq-\left[u^{c}_{1,2}\right]$ and $\left[\psi_{d}\right]=-\left[d^{c}_{j}\right]$ for all $j$. Note that the dangerous operator $\Phi_{N}\tilde{D}^{c}_{[\psi_{d}]}\bar{D}^{c}_{-[\psi_{d}]}$ would have contributed to $\bar{\theta}_{\textrm{QCD}}$ at tree-level, hence we have assumed that $\tilde{D}_{-[\psi_{d}]}$ and $\tilde{D}^{c}_{[\psi_{d}]}$ do not exist (the same discussion holds for the up type quarks). The explicit breaking parameters in the IR, $g^{u,d}_{j}$ and $\tilde{g}^{u,d}_{j}$, arise after the heavy FN states are integrated out at tree-level, and consist of a linear combination of the UV explicit breaking parameters, $\left\{\gamma\right\}$ and $\left\{\tilde{\gamma}\right\}$, respectively
\begin{align}
g^{d}_{j}&=a^{d}_{j}\cdot\gamma^{d}_{j}+b^{d}_{j}\gamma^{d}_{4}\,,\!\!\!\!\!\!\!\!\!\!\!\!\!\!\!\!\!\!\!\!\!\!\!\!\!\!\!\!\!\!\!\!&\tilde{g}^{d}_{j}&=\tilde{a}^{d}_{j}\cdot\tilde{\gamma}^{d}_{j}+\tilde{b}^{d}_{j}\tilde{\gamma}^{d}_{4}\,,\\
g^{u}_{j}&=\delta_{j3}a^{u}_{3}\gamma^{u}_{3}+b^{u}_{j}\gamma^{u}_{4}\,,\!\!\!\!\!\!\!\!\!\!\!\!\!\!\!\!\!\!\!\!\!\!\!\!\!\!\!\!\!\!\!\!&\tilde{g}^{u}_{j}&=\delta_{j3}\tilde{a}^{u}_{3}\tilde{\gamma}^{u}_{3}+\tilde{b}^{u}_{j}\tilde{\gamma}^{u}_{4}\,,
\end{align}
where the coefficients are of order unity. This simple UV completion is compatible with the selection rules of Table~\ref{table:spurion}, and the same parametric for the corrections in Eqs.~(\ref{eq:mur}-\ref{eq:dyd}) is achieved.


\medskip

\section{Backreaction Potential Scaling}\label{app:leptons}

The structure of the backreaction potential, presented in Eq.~\eqref{eq:danger} and Eq.~\eqref{eq:br}, can be understood by assuming $M^M\gtrsim M^D$ in the sterile neutrino Lagrangian, Eq.~\eqref{fnb}. In this limit, the heavy fermions, $N$, can be integrated out at tree-level, making the parametric structure of the loop corrections more transparent. While giving the correct parametric form, the approximation we present here would be quantitatively inaccurate when $M^{D}\sim M^{M}\,$, and a full numerical treatment of the 1-loop CW potential is ultimately needed. 

We use the flavor basis as it was defined in Eq.~\eqref{eq:barM^M}. Integrating out the heavy fermions leads to modifications of both the kinetic and mass terms of the lighter fermions, $N^c$, as follows
\begin{align}
\mathcal{L}_{N^c}^{\text{mass}}&=-\frac{1}{2}M^{N^{c}}_{jk}N_{j}^{c}N_{k}^{c}+\textrm{h.c.}\equiv-\frac{1}{2}\left[\left(M^{D}\right)^{T}X\right]_{jk}N_{j}^{c}N_{k}^{c}+\textrm{h.c.}\,,\label{eq:N^c Majorana}\\
{\mathcal L}_{N^c}^{\text{kin}}&=iN^{c\dagger}_{j}\overline{\sigma}^{\mu}\left[\mathbbm{1}\partial_{\mu}+X^{\dagger}X\partial_{\mu}+X^{\dagger}\left(\partial_{\mu}X\right)\right]_{jk}N^{c}_{k}\,,\label{eq:N^c Kinetic Mixing}
\end{align}
where we defined $X_{jk}=\left[\left(\overline{M}^{M}\right)^{-1}M^{D}\right]_{jk}\,$, and $\mathbbm{1}$ is the identity matrix in flavor space. Note that generically $\partial_{\mu}X\neq0$ because $X$ depends on $U_{0}$. The Higgs-independent contribution to the CW potential in Eq.~\eqref{eq:danger} is given by a quartically divergent diagram involving only insertions of Eq.~\eqref{eq:N^c Kinetic Mixing}. The scaling of the dominant contribution to $V_{\text{br}}$ in Eq.~\eqref{eq:br} can be matched to a quadratically divergent diagram involving Yukawa vertices, where a kinetic mixing insertion between $N^{c}_1$ and $N^{c}_2$ is needed to close the loop. Other (subleading) contributions to $V_{\text{br}}$ are also recovered. For example, we recover the one involving two Majorana mass insertions of Eq.~\eqref{eq:N^c Majorana}, which scales as $\sim\textrm{Tr}\left[Y^{n}Y^{n\dagger}M^{N^{c}}M^{N^{c}\dagger}\right]\,$.


\medskip

\section{About the Experimental Status of Flavor-Violating Leptonic Decays}\label{app:lepton_pheno}

Here we want to comment on the present status of searches for FV decays accompanied by missing energy in the lepton sector. We first summarize the existing searches, and then indicate some missing one that would be interesting to have from the experimental collaborations.


\medskip

\paragraph{Muon FV Decay}

FV muon decay in Eq.~\eqref{eq:muon} gives an electron/positron line with energy $E_{e}^{\text{line}}=m_\mu/2$ at the end of the SM background distribution. The current best bound on such a signal comes from the TRIUMF experiment~\cite{Jodidio:1986mz}, where $1.8\cdot 10^{7}$ $\mu^{+}$ were collected, giving the constraint $\text{BR}(\mu^{+}\!\to\! e^{+}\,\phi)< 2.6\cdot10^{-6}\,$, which translates into $f\gtrsim 2.8\cdot 10^7\text{ GeV}\,$. Since the distribution of the positrons from Eq.~\eqref{eq:muon} is isotropic, the SM background from $\mu\to e\,\nu\,\bar{\nu}$ was sensibly reduced at TRIUMF by looking only at the positrons in the direction opposite to the muon polarization. A slightly weaker bound, but less dependent on the purely axial nature of the familon couplings, was obtained from the Crystal Box detector~\cite{Bolton:1988af} by requiring an extra photon in the final state: $\text{BR}(\mu\to e\,\phi\gamma)< 1.1\cdot10^{-9}\,$, implying $f\gtrsim 4.5\cdot 10^6\text{ GeV}\,$. The latter bound can in principle be improved by the $100$ times larger luminosity of the MEG experiment~\cite{Calibbi:2017uvl}, and even larger dataset expected from MEG~II~\cite{Baldini:2018nnn}. MEG triggers are, however, optimized to improve on $\mu\to e\gamma$ only~\cite{TheMEG:2016wtm}. In principle, a dedicated analysis of electron-only events at MEG and/or MEG~II could also improve the bound from the TRIUMF experiment (see Ref.~\cite{Hirsch:2009ee} for a similar discussion in the context of R-parity violating SUSY models). In principle, also the Mu3e experiment can have some sensitivity to single electron events if an electron trigger is developed together with a detector upgrade (see fore example Ref.~\cite{talk}).\footnote{We thank Lorenzo Calibbi for many discussions about muon beam experiments.}


\medskip

\paragraph{Tau FV Decay}

A dedicated analysis of ARGUS data, based on an integrated luminosity of $472\text{ pb}^{-1}$~\cite{Albrecht:1995ht,Igonkina:1996ar}, sets a bound on the FV tau decays in Eq.~\eqref{eq:tau}. From $\text{BR}(\tau\!\to\!\phi\,\,e\,,\mu)< 2.6\,,4.5\cdot10^{-3}\,$, we get $f\gtrsim 2.5\,,1.8\cdot 10^5\text{ GeV}\,$. An improvement of the bound on the FV tau decay with a muon in the final state has been recently obtained from Belle data~\cite{Yoshinobu:2017jti}, with an integrated luminosity of $1020\text{ pb}^{-1}$: $\text{BR}(\tau\!\to\!\phi\,\,\mu)< 1.1\cdot10^{-4}\,$, corresponding to $f\gtrsim1.2\cdot 10^6\text{ GeV}\,$. Future B-factories, like Belle~II, will produce $10^{11}$ $\tau$'s, pushing the reach on $f$ to $4\cdot 10^6\text{ GeV}\,$, where we have assumed the systematic uncertainties to be under control so that the bound on $\text{BR}(\tau\to \mu\,\phi)$ scales as the squared root of the luminosity. It would be interesting to probe also $\text{BR}(\tau\to \mu\,\phi\, \gamma)\,$, for which the SM background is sensibly reduced (see Ref.~\cite{Heeck:2016xkh} for a similar comment in the context of light $Z'$).


\medskip
\bibliography{hierarchion}

\end{document}